\documentclass[bjps]{imsart}

\RequirePackage{amsthm,amsmath,amsfonts,amssymb}
\RequirePackage[authoryear]{natbib}
\RequirePackage[colorlinks,citecolor=blue,urlcolor=blue]{hyperref}
\RequirePackage{graphicx}

\usepackage{placeins}
\usepackage{graphicx}%
\usepackage{multirow}%
\usepackage{amsthm}%
\usepackage{mathrsfs}%
\usepackage[title]{appendix}%
\usepackage{xcolor}%
\usepackage{textcomp}%
\usepackage{manyfoot}%
\usepackage{booktabs}%
\usepackage{algorithm}%
\usepackage{algorithmicx}%
\usepackage{algpseudocode}%
\usepackage{listings}%
 \usepackage{float}
\usepackage{bm}
\usepackage{subfigure}

\startlocaldefs
\theoremstyle{plain}

\newtheorem{theorem}{Theorem}[section]
\newtheorem{remark}{Remark}[section]
\newtheorem{proposition}{Proposition}[section]
\newtheorem{lemma}[theorem]{Lemma}
\theoremstyle{definition}

\endlocaldefs

\begin{document}
\begin{frontmatter}

\title{Integral Fractional Ornstein-Uhlenbeck Process Model for Animal Movement}
\runtitle{Integral fOU Model for Animal Movement}

\begin{aug}
\author[A]{\inits{J.H.}\fnms{Jos\'e Hermenegildo}~\snm{Ram\'irez Gonz\'alez}\thanksref{t1}\ead[label=e1]{hermenegildo.ramirez@cimat.mx}},
\author[A]{\inits{Y.}\fnms{Ying}~\snm{Sun}\thanksref{t1}\ead[label=e2]{ying.sun@kaust.edu.sa}}

\thankstext{t1}{These authors contributed equally to this work.}

\address[A]{CEMSE Division, Statistics Program, King Abdullah University of Science and Technology (KAUST),
Thuwal, 23955-6900, Makkah, Saudi Arabia\printead[presep={,\ }]{e1,e2}}
\end{aug}

\begin{abstract}
The study of animal telemetry is crucial in ecology, providing valuable information on movement patterns, behavior, and habitat use of various species, which is essential for conservation and management efforts. Numerous models in the literature address animal telemetry by modeling velocity, telemetry data itself, or both processes jointly through a Markovian approach. In this work, we propose a novel approach by modeling the velocity of each coordinate axis for animal telemetry data using a fractional Ornstein-Uhlenbeck (fOU) process. The integral of the fOU process models the position data in animal telemetry. This proposed model is particularly flexible in capturing long-range memory effects. The Hurst parameter $H \in (0,1)$ plays a crucial role in the integral fOU process, determining the long-range memory. The integral fOU process is nonstationary; a higher Hurst parameter ($H > 0.5$) indicates stronger memory, leading to trajectories with transient tendencies, while a lower Hurst parameter ($H < 0.5$) implies weaker memory, resulting in trajectories with recurring trends. When $H = 0.5$, the process reduces to a standard integral Ornstein-Uhlenbeck process. We develop a simulation algorithm for telemetry trajectories using finite-dimensional distributions and employ the maximum likelihood method for parameter estimation, with its performance evaluated through simulation studies. Finally, we present a telemetry application involving Fin Whales dispersing throughout the Gulf of California.
\end{abstract}


\end{frontmatter}

\section{Introduction}\label{sec:intro}

Our interest in animal movements is driven by scientific inquiry and the need to inform decisions related to the management and conservation of natural resources. Leveraging telemetry data has yielded valuable insights, allowing researchers to investigate essential ecological hypotheses concerning space utilization, such as the animal's location, its journey, and its preferred habitat (\cite{kays2015terrestrial, hebblewhite2010distinguishing, nathan2008movement, patterson2008state}).

Population dynamics encompass processes like births, deaths, immigration, and emigration, which are fundamental to understanding species persistence and adaptation. One of the most relevant techniques for animal conservation is the study of telemetry combined with statistical models. The application of statistical models to telemetry data has become an indispensable tool for understanding animal behavior, movement patterns, and population dynamics \citep{patterson2008state, johnson2008modeling, schick2008understanding}. These models uncover hidden patterns and relationships within the data, allowing for informed conservation efforts \citep{cagnacci2010animal, forester2009using, boyce2002relative}. By integrating telemetry data into statistical frameworks, researchers can make scientifically grounded predictions and guide effective strategies for wildlife conservation and resource management.



In this work, we propose modeling animal telemetry using the fractional Ornstein–Uhlenbeck (fOU) process. The Ornstein–Uhlenbeck (OU) process is widely utilized across various scientific fields due to its properties such as mean reversion, long-term stationarity, and exponentially decaying correlations. In environmental sciences, a Gamma extension of the OU process has been used to analyze earthquake magnitudes \citep{HABTEMICAEL2014147}. In biology, it is employed for phenotypic trait evolution, considering factors like adaptation and migration \citep{BARTOSZEK201735}. In \citep{BUONOCORE201559}, the OU process is applied to model the membrane potential dynamics of single neurons. The study constructs inhomogeneous leaky integrate-and-fire stochastic models using restricted Gauss–Markov processes, which include the OU process. This modeling approach analyzes neuron activity in the presence of a lower reflecting boundary and periodic input signals, focusing on the first-passage time problem through a time-dependent threshold. In \cite{Gao}, the OU process is used to model the influence of environmental variability on nutrient and plankton dynamics. This modeling approach helps analyze conditions under which plankton populations either persist or become extinct due to fluctuating environmental factors. The study provides valuable insights into the stability and resilience of marine ecosystems under varying environmental conditions. Below, we mention the definition of the OU process and relevant applications of this process in animal telemetry.

We employ a movement model based on animal telemetry velocity, building on previous methodologies \cite{John, Hooten, Tor}. Gaussian processes (GPs) have been applied in these works to model animal movement velocity within hierarchical models. Specifically, the velocity is modeled through linear stochastic differential equations (SDE), which can be expressed as GPs with a suitable covariance structure \cite{Sarkka}. Notably, all random walk movement models that can be formulated as linear SDEs are also equivalent to GPs \cite{Hooten, Tor}. Among the crucial models for velocity telemetry, the correlated random walk (CRW) model for animal movement can be formulated in both discrete-time (\cite{McClintock}) and continuous-time \cite{gurarie}. In the continuous-time version, a correlated velocity model is employed, also known as an Ornstein-Uhlenbeck velocity model or integrated Ornstein-Uhlenbeck model \cite{John}. Many animals exhibit repeated return patterns to key locations, a common feature in sedentary populations. This behavior induces structured temporal dependence that is not well captured by  processes lacking long-range dependence. Specifically, we seek to develop a covariance matrix that represents the correlation structure in positional observations of an animal undergoing an autocorrelated continuous-time GP. Our starting point is an Ornstein–Uhlenbeck velocity model for the animal’s movement, described by the following equations:
\begin{equation*}
  d\mu(t)=v(t)dt  \text{ and }dv(t)=-a(t)v(t)dt+b(t)dW(t),
\end{equation*}
where $v$ is its velocity, $W(t)$ is a Brownian motion, and $a(t)$ and $b(t)$ are time-varying coefficients. In the case of constant parameters ($a(t)=a$ and $b(t)=b$) the covariance function of the Ornstein–Uhlenbeck (OU) process is well known \cite{Gardiner}. When we adopt the initial condition $v(0)=\sigma \int_{-\infty}^{0} e^{\beta u}\,dW(u)$, its covariance is equivalent to the exponential covariance, i.e.,
\begin{equation}\label{vcov}
    \text{cov}(v(t),v(s))=\frac{b^2}{2a}\exp(-a|t-s|),
\end{equation}
and the process is therefore stationary.

To relate the covariance of the velocity process to that of the position process $\mu$, with initial condition $\mu(0)$ and $\mu(t)=\mu(0)+\int_0^t v(u)\,du$, its covariance function is
\begin{equation}\label{cove}
  \text{cov}(\mu(t),\mu(s))=\int_0^t\int_0^s \text{cov}(v(u),v(w))\,du\,dw .
\end{equation}

\cite{Tor} consider the Matérn covariance function in the equation (\ref{vcov}) and they studied the telemetry of the animals by means of the covariance function given in (\ref{cove}).  

On the other hand, existing models do not allow us to study long-range dependence. Incorporating long-range dependency into movement models is important for achieving a more accurate and realistic representation of temporal dependencies, leading to improved forecasting and overall model performance. One way to introduce long-range dependence is by incorporating more general noise \( W_H(t) \) through a fractional Ornstein–Uhlenbeck (fOU) process \cite{fBM1} to model the velocity of each trajectory. Specifically,
\begin{equation}
  dv_H(t)=-\beta v_H(t)\,dt+\sigma\,dW_H(t), \qquad d\mu_H(t)=v_H(t)\,dt,
\end{equation}
where $v_H$ is an fOU process which extends the classical Ornstein–Uhlenbeck process and has important applications in finance, physics, and ecology. In addition, with the initial condition $v_H(0)=\sigma \int_{-\infty}^{0} e^{\beta u}\,dW_H(u)$, $v_H(t)$ is stationary. Unlike the traditional Ornstein-Uhlenbeck process, which has short-term memory, the fOU process exhibits memory effects on a longer time scale, which is called the long-memory property. This implies that the past values of the process have a significant influence on future values. The Hurst parameter $H\in (0,1)$ is a crucial parameter in fOU process, as it determines the degree of long-range dependence or memory. A higher Hurst parameter ($H > 0.5$) indicates stronger memory, while a lower Hurst parameter ($H < 0.5$) implies weaker memory. When $H = 0.5$, the process is reduced to a standard Ornstein-Uhlenbeck process. The Hurst parameter $H<0.5$ indicates negative correlation between increments, resulting in trajectories characterized by recurrence. Conversely, $H>0.5$ indicates positive correlation between increments, resulting in trajectories that are transient. When $H=0.5$, the increments are uncorrelated. Estimating model parameters of a fOU process is challenging. We can find several references \cite{Alexandre, Nualart, Tanaka, Xiao} for the estimation of those parameters and asymptotic laws but the asymptotic results require the derivative of the position process. However, telemetry data usually do not include the velocity associated with the trajectory, and telemetry data are typically sparse in time, so that numerical methods often fail to estimate the derivative accurately. Therefore, we resort to the finite-dimensional laws of the position process, and the estimation is carried out via likelihood. 

The rest of the paper is organized as follows. In Section \ref{s2}, we present the definitions of the velocity and position process for modelling animal movement, We establish some theoretical results for the position process $\mu_H(t)$, including the derivation of its finite-dimensional distributions.
 In Section \ref{sec:simulation}, we provide algorithms for simulating and predicting the position process $\mu_H(t)$. In Section \ref{sec:3.2} we derive explicitly likelihood function to estimate the parameters associated with the position process. In Section \ref{practical} we show the detailed analysis of telemetry data of Fin Whales whose habitat is the Gulf of California. Finally, in Section \ref{sec:discussion} we conclude with a general discussion of the model and methodology. All the code is available on GitHub: \url{https://github.com/joseramirezgonzalez/Integral_fractional_Brownian_motion}.

\section{Integral Fractional Ornstein-Uhlenbeck Process}\label{s2}
In this section, we first define the velocity and position process of animal movement. We will then provide examples of trajectories to visualize features of the process. Finally, we develop the finite-dimensional distributions of the position process to estimate the parameters.  
Let $(W_{H}(t))_{t\geq 0}$ be a fractional Brownian motion (fBm), represented by a centered Gaussian process with a covariance~function
\begin{equation}
    c_{H}(s,t):=\frac{1}{2} \Bigg(t^{2H}+s^{2H}-|t-s|^{2H}\Bigg), \text{ where } H \in (0,1) \text{ is the Hurst parameter}.
\end{equation}   
 Subsequently, \cite{fBM1} introduced the fractional Ornstein-Uhlenbeck (fOU) process $v_H(t)$, defined by the following SDE:
\begin{equation}\label{lh}
    dv_H(t)=-\beta v_{H}(t)dt+\sigma dW_{H}(t), \text{ where } \beta, \sigma > 0 \text{ are constants.}
\end{equation}
We propose to model the velocity of animal movement, $v_H(t)$, as a fOU process, and the position of the movement is given by the integral of the velocity. That is:
\begin{equation}\label{mpr}
    \mu_{H}(t)=\mu_H(0)+\int_{0}^{t}v_H(s)ds.
\end{equation}
To study animal telemetry, we consider two independent positions, $\mu_{H_1}(t)$ and $\mu_{H_2}(t)$, defined as in (\ref{mpr}), each representing the position of the animal in longitude and latitude.

\subsection{Theoretical Properties}\label{sec:theory}
In this section, we study certain properties of the processes defined in (\ref{lh}) and (\ref{mpr}). In Proposition \ref{PGH}, we show that the velocity and position processes are Gaussian processes, and we derive the explicit forms of their mean and covariance functions. The proof can be found in Appendix A.

\begin{proposition}\label{PGH}
    $\mu_H(t)$ is a Gaussian process with mean 
\begin{equation}
    m(t)=\mu_{H}(0)+v_{H}(0)\left(\frac{1-e^{-\beta t}}{\beta}\right)
\end{equation}    
and covariance function 
\begin{equation}\label{QMH}
    Q_{H,\beta,\sigma^{2}}(s,t):=\sigma^{2}\int_{0}^{t}\int_{0}^{s}e^{-\beta v}\left(c_{H}(t-v,s-u)\right)e^{-\beta u}dudv
\end{equation}
\end{proposition}

\begin{figure}[H]
\centering
\subfigure[$(H=0.25,\sigma=3,\beta=5)$]{\includegraphics[width=42mm]{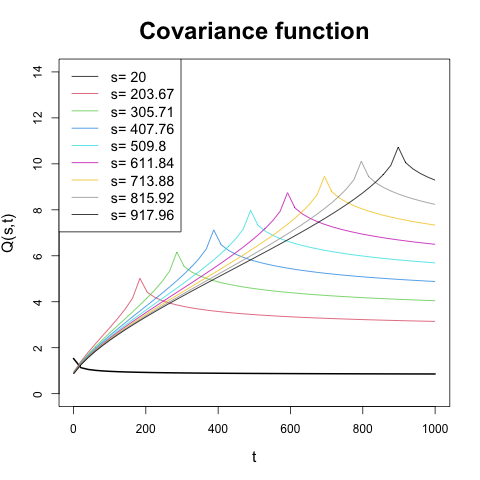}}
\subfigure[$(H=0.25,\sigma=3,\beta=0.1)$]{\includegraphics[width=42mm]{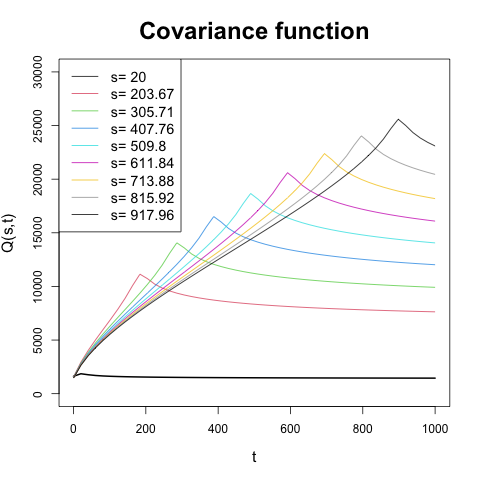}}
\subfigure[$(H=0.25,\sigma=3,\beta=30)$]{\includegraphics[width=42mm]{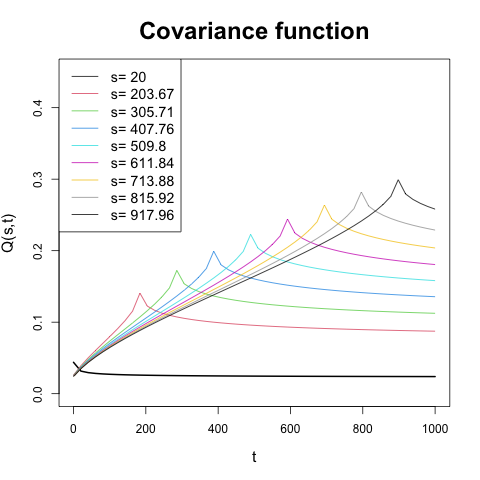}}\vspace{3mm}
\subfigure[$(H=0.5,\sigma=3,\beta=5)$]{\includegraphics[width=42mm]{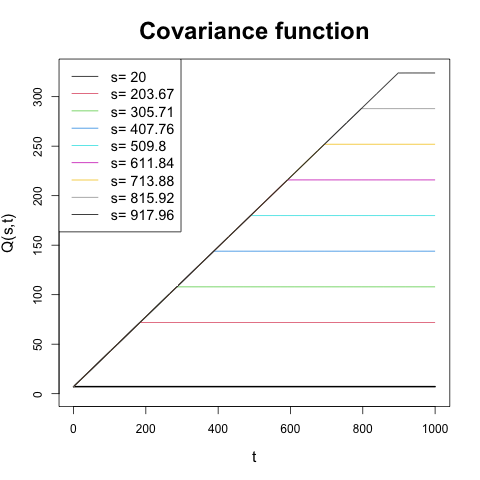}}
\subfigure[$(H=0.5,\sigma=3,\beta=0.1)$]{\includegraphics[width=42mm]{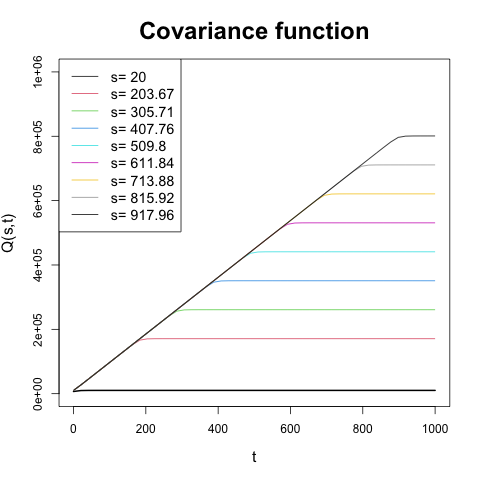}}
\subfigure[$(H=0.5,\sigma=3,\beta=30)$]{\includegraphics[width=42mm]{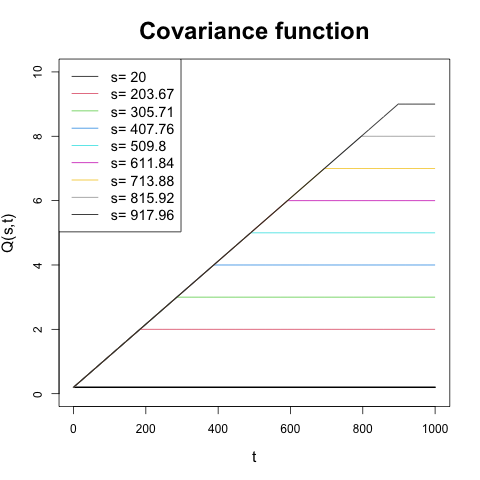}}\vspace{3mm}
\subfigure[$(H=0.75,\sigma=3,\beta=5)$]{\includegraphics[width=42mm]{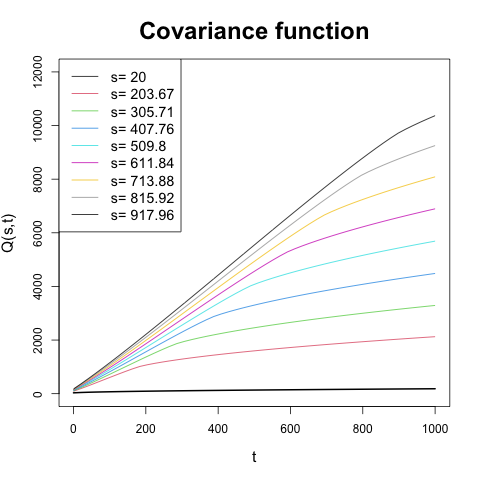}}
\subfigure[$(H=0.75,\sigma=3,\beta=0.1)$]{\includegraphics[width=42mm]{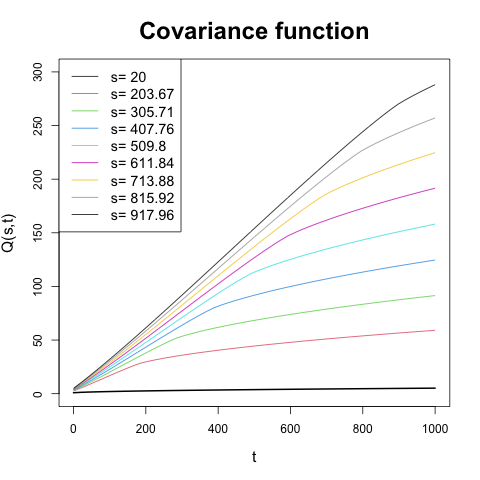}}
\subfigure[$(H=0.75,\sigma=3,\beta=30)$]{\includegraphics[width=42mm]{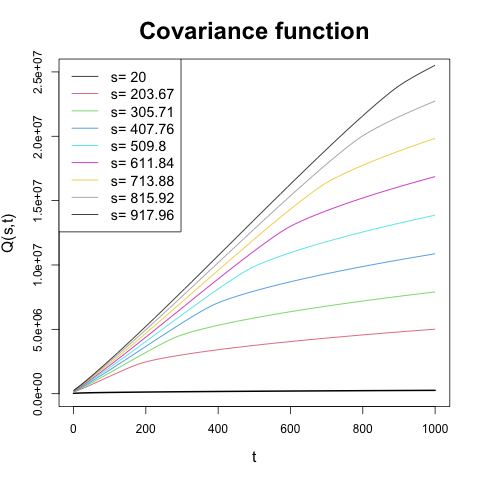}}
\caption{Illustrative Examples of Covariance Functions \( Q_{H,\beta,\sigma^{2}}(s,t) \): These plots show the covariance functions associated with the positional process \( \mu_H(t) \), highlighting the effects of parameters \( H \), \( \beta \), and \( \sigma^2 \) on temporal correlations.}\label{FCova}
\end{figure}

Figure \ref{FCova} shows examples of covariance functions given by equation (\ref{QMH}). In the fBm with covariance function $c_H$, a Hurst parameter $H<0.5$ indicates negative correlation between increments, resulting in recurrent trajectories. Conversely, $H>0.5$ indicates positive correlation between increments, leading to transient trajectories. When $H = 0.5$, the increments are uncorrelated. Consequently, for fixed $s$, when $H>0.5$ ((g), (h), and (i)), the covariance function $Q_{H,\beta,\sigma^{2}}(s,\cdot)$ is increasing over $(0,\infty)$. When $H<0.5$ ((a), (b), and (c)), $Q_{H,\beta,\sigma^{2}}(s,\cdot)$ increases over $(0,s)$ and decreases over $(s,\infty)$. Finally, when $H=0.5$ ((d), (e), and (f)), $Q_{H,\beta,\sigma^{2}}(s,\cdot)$ increases over $(0,s)$ and is constant over $(s,\infty)$.

In Proposition~\ref{Prop3.2}, we derive a recursive equation for the position and velocity vector. Although the process is not autoregressive in the strict sense when $H \neq \frac{1}{2}$ due to temporal correlation in the noise, the expression retains an autoregressive structure. The proof can be found in Appendix~A.

\begin{proposition}\label{Prop3.2}
Let $\alpha_{H}(t)=(\mu_{H}(t),v_{H}(t))$. For any $\Delta>0$ we have the equality $\alpha(t+\Delta)=T\alpha(t)+\eta_{t,\Delta}$. Here, 
\begin{equation*}
T=\begin{pmatrix}
1 & \frac{1-e^{-\beta \Delta}}{\beta} \\
0 & e^{-\beta \Delta} 
\end{pmatrix};
\end{equation*}
and $\eta_{t,\Delta} \sim N((0,0),\Sigma_{t,\Delta})$ where
$$\eta_{t,\Delta}=\Bigg(\sigma\int_{t}^{t+\Delta}e^{\beta(s-(t+\Delta))}dW_{H}(s),\sigma\int_{t}^{t+\Delta}\frac{1-e^{\beta(s-(t+\Delta))}}{\beta}dW_{H}(s)\Bigg)^{t}$$
$\text{ and }\Sigma_{t,\Delta}:=\text{cov}(\eta_{t,\Delta},\eta_{t,\Delta}).$ 
\end{proposition}

\begin{remark}
Let $\Delta_i:=t_{i+1}-t_i$, $i=0,1,...,n-1$  and $\eta_{i}:=\eta_{t_i,\Delta_i}$. When $H=\frac{1}{2}$ we will have that $\eta_{t_i,\Delta_i}$ and $\eta_{t_j,\Delta_j}$ are independent and the stochastic process $(\mu_H(t), v_H(t))$ is Markovian (see \cite{John}). However, in the general case $H\neq \frac{1}{2}$ it is easy to see that $\eta_{t_i,\Delta_i}$ and $\eta_{t_j,\Delta_j}$ are correlated for $i \neq j$. In particular, to simulate finite-dimensional distributions of $\mu_H(t)$ we have to jointly simulate the random vectors $(\eta_{t_0,\Delta_0},...,\eta_{t_{n-1}, \Delta_{n-1}})$. In Section \ref{sec:simulation} we give an efficient method of performing trajectory simulations of $\mu_H(t)$ by means of finite-dimensional distributions. Another consequence of such correlation is that that we cannot make use of the trajectory estimation technique given in (\cite{John}) Kalman-Filter method.
\end{remark}

The recursive structure derived in Proposition~\ref{Prop3.2} also serves as a foundation for the following result. Predicting the velocity process $v_{H}(t)$ given an observed trajectory necessitates computing the covariance between the velocity and position $\mu_{H}(t)$ processes, which can be non-trivial. We suggest obtaining predictions using the auxiliary variables defined in the following Proposition, the proof can be found in Appendix A.

\begin{proposition}\label{Propv}
We have recurring equations between velocity, position and noise. This is:
\begin{equation}\label{v1}
    v_{H}(t_i+\Delta_i)-e^{-\beta \Delta}v_{H}(t_i)=\zeta_{t_i}^{(1)}+\zeta_{t_i}^{(2)}.
\end{equation}
Here,
\begin{equation}\label{z1}
  \zeta^{(1)}_{t_i}:=\sigma (W_{H}(t_{i}+\Delta_i)-e^{-\beta \Delta_i}W_{H}(t_i)), \text{ for }  i=0,....,n-1
\end{equation}
and
 \begin{equation}\label{sv1}
    \begin{split}
       \zeta^{(2)}_{t_i}&=-\sigma\beta\int_{t_i}^{t_i+\Delta_i}e^{\beta(s-(t_i+\Delta_i))}W_{H}(s)ds\\
       &=\beta (1-e^{-\beta \Delta_i})\mu_{H}(0)-\beta( \mu_{H}(t_i+\Delta_i)-e^{-\beta \Delta_i}\mu_{H}(t_i)),
    \end{split}
\end{equation}
\end{proposition}

We now analyze the asymptotic behavior of the process \( \mu_H(t) \) as the parameter \( \beta \) tends to infinity. The following proposition characterizes the limiting behavior of its finite-dimensional distributions.
\begin{proposition}\label{cov_beta}
     Since \( \mu_H(t) \) is Gaussian and centered for each \( \beta > 0 \), the convergence of its covariance function to zero as \( \beta \to \infty \) implies that
\[
\lim_{\beta \to \infty} \mu_H(t) \to \mu_H(0) \quad \text{in the sense of finite-dimensional distributions}.
\]
\end{proposition}

\begin{remark}\label{Remarkbeta}
Proposition~\ref{cov_beta} suggests that, when the true value of \( \beta \) is large, likelihood-based inference methods may face limitations due to a potential loss of sensitivity of the likelihood function with respect to \( \beta \). 
\end{remark}

\section{Simulation and Predictions}\label{sec:simulation}
In this section we provide a method to simulate the finite-dimensional distributions of $\mu_H(t)$. Given the continuity of the trajectories of $\mu_H(t)$ we show that the approximation via finite-dimensional distributions converges uniformly.
\subsection{Simulation method}\label{sec:simulation1}
We proceed to simulate the finite-dimensional distributions of $\mu_{H}(t)$. Given the result of Proposition \ref{PGH}, our objective is to simulate a Gaussian vector with the covariance matrix:
\begin{equation}\label{Qr}
     Q_{t_i,t_j}:=\sigma^{2}\int_{0}^{t_i}\int_{0}^{t_j}e^{- \beta v}\Bigg(c_{H}(t_i- v,t_j-u)\Bigg)e^{- \beta u}dudv
\end{equation}
for $\bm{t}=(t_1,...,t_n)$, where $0=t_0< t_1 <t_2<...<t_n$ with $n\in \mathbb{N}$. 

\begin{remark}\label{Integral}
   For numerical calculations of double integrals, the size of the integration region significantly impacts the efficiency and accuracy of numerical integration techniques like quadrature rules. Using smaller regions helps to lower computational complexity and minimize errors. 
\end{remark}

For the simulation of the finite-dimensional distributions of $\mu_H(\bm{t})$, we need to define some auxiliary variables, which will allow us to perform the simulations more efficiently. Let $\Delta_i:=t_{i+1}-t_i$, $i=0,1,...,n-1$, and assume that $v_H(0)=0$. We define 
\begin{equation}\label{Z^2}
    y_{i}:=-\frac{\zeta_{t_{i}}^{(2)}}{\sigma\beta}:=\int_{t_{i}}^{t_i+\Delta_i}e^{\beta(s-(t_i+\Delta_i))}W_{H}(s)ds \text{ for }  i=0,....,n-1
\end{equation}
and $\mu_i:=\mu_{H}(t_i)$ for $i=1,...,n$ and we suppose that we know $\mu_0:=\mu_H(0)$. 
In Proposition~\ref{Propy}, we explicitly give the finite-dimensional distributions of $\mu_H(t)$. The proof can be found in Appendix~A.

\begin{proposition}\label{Propy} We have that 
\begin{equation}\label{distribution}
    (\mu_1,...,\mu_n)^{t}=(\mu_0,...,\mu_0)^{t}+\Sigma_{\beta,\sigma,(t_0,...,t_n)}\cdot(y_0,...,y_{n-1})^{t}\text{ where } (y_0,...,y_{n-1})^{t}\sim N(0,\Sigma_{H,\beta}).
\end{equation}
Therefore, $(\mu_1,...,\mu_n)^{t}\sim N((\mu_0,...,\mu_0)^{t},\Sigma_{\beta,\sigma,(t_0,...,t_n)} \Sigma_{H,\beta}\Sigma_{\beta,\sigma,(t_0,...,t_n)}^{t})$, where $\Sigma_{\beta,\sigma,(t_0,...,t_n)}$ is a lower triangular matrix with $\sigma$ as a scaling factor and exponential terms $e^{-\beta(t_i-t_j)}$ that account for the temporal correlations between different time points and for $i,j=0,...,n-1$ we have that:
\begin{equation}\label{SigmaC}
\Sigma_{H,\beta}(i+1,j+1):=
 \int_{t_j}^{t_j+\Delta_j}\int_{t_i}^{t_i+\Delta_i}e^{\beta(u-(t_i+\Delta_i))}e^{\beta(v-(t_j+\Delta_j))}c_{H}(u,v)dudv, 
\end{equation}
\end{proposition}

According to Remark \ref{Integral}, it will be numerically more efficient to compute the finite-dimensional distributions of $\mu_H(t)$ using equation (\ref{distribution}) rather than directly from equation (\ref{Qr}).

Now, letting $\Delta_i=\Delta$, we have a way to perform simulations efficiently as shown below.
\begin{remark} (Fast simulations) For compute $\Sigma_{H,\beta}(i,j)$. In general case, we need to compute $\frac{n(n-1)}{2}$ double integral. But, if we assume that $\Delta_i=\Delta$. We only need compute $n$ double integral and $n$ simple integral. Suppose that $j\leq i$, it is easy to check that
\begin{equation*}
\begin{split}
\Sigma_{H,\beta}(i+1,j+1)&=\frac{1-e^{-\beta \Delta}}{2\beta}\Bigg(\int_{0}^{\Delta}e^{-\beta u}(t_j+\Delta-u)^{2H}du+\int_{0}^{\Delta}e^{-\beta u}(t_i+\Delta-u)^{2H}du\Bigg)\\
 &-\frac{1}{2}\int_{0}^{\Delta}\int_{0}^{\Delta}e^{-\beta u}e^{-\beta v}|v-u+(i-j)\Delta|^{2H}dudv. 
\end{split}
\end{equation*}
Let
\begin{equation}\label{sima3}
\begin{split}
h(i)&:=\int_{0}^{\Delta}e^{-\beta u}((i+1)\Delta-u)^{2H}du, \phantom{|}k(i)\\
&:=\int_{0}^{\Delta}\int_{0}^{\Delta}e^{-\beta u}e^{-\beta v}|v-u+i\Delta|^{2H}dudv,\phantom{|}i=0,....,n-1.
\end{split}
\end{equation}
Then, for $j\leq i$
\begin{equation}\label{sima2}
   \Sigma_{H,\beta}(j+1,i+1)=\Sigma_{H,\beta}(i+1,j+1)=\frac{1-e^{-\beta \Delta}}{2\beta}\Bigg(h(i)+h(j)\Bigg)-\frac{1}{2}k(i-j)
\end{equation}
and $\Sigma_{\beta,\sigma}$ is a lower triangular matrix with $\Sigma_{\beta,\sigma}(i,j)=e^{(i-j)\beta\Delta}$.
\end{remark}

Therefore, we can use Proposition \ref{Propy} to simulate the finite-dimensional distributions of $\mu_H(\bm{t})$. These simulations can be performed as described in Algorithm~\ref{alg:pos}.
\begin{algorithm}[H]
\caption{Simulation of the finite-dimensional distributions of the position process $\mu_{H}(\bm{t})$ on the time set $\bm{t}=(t_1,...,t_n)$.}\label{alg:pos}
\begin{algorithmic}
\State 1.- Perform simulations of the auxiliary variables $(y_0,...,y_{n-1})^{t}$ according to Proposition \ref{Propy} and equations (\ref{SigmaC}) or (\ref{sima2}), as appropriate.
\State 2.- Compute the lower triangular matrix $\Sigma_{\beta,\sigma}$ which is given by $\Sigma_{\beta,\sigma}(i,j)=\sigma e^{(t_i-t_j)\beta}$.
\State 3.-Compute $(\mu_1,...,\mu_n)^t$ according to equation (\ref{distribution}).
\State 4.- Return the simulation of the finite-dimensional distributions of the positional process $\mu_{H}(\bm{t})$.
\end{algorithmic}
\end{algorithm}

Based on Proposition~\ref{Propv}, we provide an algorithm to simulate the velocity process $v_H(t)$ given an observed trajectory and the corresponding maximum likelihood estimators. 

To apply the theory of conditional Gaussian distributions, it is enough to calculate the covariance between $(\zeta _{t_j}^{(1)},\zeta _{t_i}^{(1)})$, $(\zeta _{ t_j}^ {(1)},\zeta _{t_i}^{(2) })$ and $(\zeta _{t_j }^{(2)},\zeta _{t_i}^{(2)}) $. 
 It is easy to check that:
\begin{equation}\label{v2}
\begin{split}
\text{cov}\Bigg(\zeta_{t_j}^{(1)},\zeta_{t_i}^{(1)}\Bigg)
    =\sigma^{2}&\Bigg[c_{H}(t_j+\Delta_j,t_i+\Delta_i)-e^{-\beta \Delta_i}c_{H}(t_j+\Delta_j,t_i)\\
    &-e^{-\beta \Delta_j}c_{H}(t_i+\Delta_i,t_j)+e^{-\beta (\Delta_i+\Delta_j)}c_{H}(t_i,t_j)\Bigg]
\end{split}
\end{equation}
\begin{equation}\label{v3}
\begin{split}
\text{cov}\Bigg(\zeta_{t_j}^{(1)},\zeta_{t_i}^{(2)}\Bigg)
    &=-\sigma^{2}\beta\Bigg[e^{-\beta \Delta_i}\int_{t_i}^{t_i+\Delta_i}e^{\beta(s-t_i)}c_{H}(t_j+\Delta_j,s)ds\\
    &-e^{-\beta( \Delta_i+\Delta_j)}\int_{t_i}^{t_i+\Delta_i}e^{\beta(s-t_i)}c_{H}(t_j,s)ds\Bigg]
\end{split}
\end{equation}
\begin{equation}\label{v5}
\begin{split}
\text{cov}\Bigg(\zeta_{t_j}^{(2)},\zeta_{t_i}^{(2)}\Bigg)
&=\sigma^{2}\beta^{2}e^{-\beta( \Delta_i+\Delta_j)}\int_{t_j}^{t_j+\Delta_j}\int_{t_i}^{t_i+\Delta_i}e^{\beta(v-t_j)}e^{\beta(u-t_i)}c_{H}(u,v)dudv
\end{split}
\end{equation}


To predicting the velocity process $v_{H}(t)$ given a observed trajectory and the maximum likelihood estimators (MLEs) $\hat{\sigma}$, $\hat{\beta}$ and $\hat{H}$ (for MLEs see Section \ref{sec:3.2}) can be accomplished using Algorithm~\ref{alg:velocity}.

\begin{algorithm}[H]
\caption{Predict the velocity process $v_{H}(\bm{t})$ given an observed trajectory on time set $\bm{t}=(t_1,...,t_n)$.}\label{alg:velocity}
\begin{algorithmic}
\State 1.- Compute the auxiliary variables $\zeta_{t_i}^{(2)}$ for $i=0,...,n-1$ defined in Equation (\ref{sv1}) based on the observed trajectory and the maximum likelihood estimator $\hat{\beta}$.
\State 2.- Simulate a conditional realization of the auxiliary variables $\Bigg(\zeta_{t_0}^{(1)},...,\zeta_{t_{n-1}}^{(1)}\Bigg) \Bigg| \Bigg(\zeta_{t_0}^{(2)},...,\zeta_{t_{n-1}}^{(2)}\Bigg)$, using the covariance provided in Equations (\ref{v2}), (\ref{v3}) and (\ref{v5}), and the MLE $\hat{\sigma}$, $\hat{\beta}$ and $\hat{H}$.
\State 3.- Set initial velocity: $v_{H}(0) = 0$.
\State 4.- Recursively obtain the velocity predictions $v_{H}(\bm{t})$ using Equation (\ref{v1}), the MLE $\hat{\beta}$, and the generated values of the auxiliary variables $\zeta_{t_i}^{(1)}$ and $\zeta_{t_i}^{(2)}$ for $i=0,...,n-1$. 
\State 5.- Return the velocity predictions $v_{H}(\bm{t})$. 
\end{algorithmic}
\end{algorithm}

\begin{remark}
In Algorithm~\ref{alg:velocity}, it is not necessary to assume $v_H(0) = 0$. In the general case, it is sufficient to consider:
\begin{equation}
    \begin{split}
       \zeta^{(2)}_{t_i} &= -\sigma\beta\int_{t_i}^{t_i+\Delta_i}e^{\beta(s-(t_i+\Delta_i))}W_{H}(s)\,ds \\
       &= \beta(1 - e^{-\beta \Delta_i})\mu_{H}(0) + v_H(0)(1 - e^{-\beta \Delta_i}) - \beta\left( \mu_H(t_i+\Delta_i) - e^{-\beta \Delta_i}\mu_H(t_i) \right),
    \end{split}
\end{equation}
which incorporates the initial velocity $v_H(0)$ into the expression for $\zeta^{(2)}_{t_i}$.
\end{remark}

Finally, the position process $\mu_{H}(t)$ can be predicted from an observed trajectory and the maximum likelihood estimates $\hat{\sigma}$, $\hat{\beta}$, and $\hat{H}$, as described in Algorithm~\ref{alg:position}.

\begin{algorithm}
\caption{Let ($m \in \mathbb{N}$). Predict $\mu_{H}(\bm{t})$ for the next $m$ steps given an observed trajectory on time set $\bm{t}=(t_1,...,t_n)$.}\label{alg:position}
\begin{algorithmic}
\State 1.- Compute the auxiliary variables $y_i = \frac{1}{\sigma} \left((e^{-\beta \Delta_i} - 1)\mu_0 - e^{-\beta \Delta_i}\mu_i + \mu_{i+1}\right)$ for $i=0,...,n-~1$. based on the observed trajectory and the maximum likelihood estimators $\hat{\beta}$ and $\hat{\sigma}$.
\State 2.- Using Gaussian conditional distributions and the covariance matrix $\Sigma_{H,\beta}$, Simulate  $y_i$ for $i=n, n+1, ..., n+(m-1)$.
\State 3.- Compute the predictions $\mu_{i}$ for $i=n+1,...,n+m$ using the equation $\mu_{i+1} = (1 - e^{-\beta \Delta_i})\mu_0 + e^{-\beta \Delta_i}\mu_i + \sigma y_i$ \text{ for } i=n, ..., n+(m-1).  
\end{algorithmic}
\end{algorithm}

\section{Inference}\label{sec:3.2}
In the literature, such as \cite{Alexandre, Nualart, Tanaka, Xiao}, we can find theoretical results for various estimators of the parameters $(\sigma,\beta, H)$ . However, these results require knowing the velocity trajectory or a specific sample of it. In practice, it is common that only sparsely sampled position trajectories are available, so we resort to maximum likelihood estimators to estimate the parameters.

First, we give some properties of the likelihood function associated with the finite-dimensional distributions of $\mu_H(t)$. Then, we describe a method to obtain Maximum Likelihood Estimators (MLEs). From the equation (\ref{distribution}) we know that $(\mu_1-\mu_0,...,\mu_n-\mu_0)\sim N(0,\Sigma_{\beta,\sigma}\Sigma_{H, \beta}\Sigma_{\beta,\sigma}^{t})$. Then, the log-likelihood function is given by:
\begin{equation}
\begin{split}
    \log&\Bigg(L\Bigg((\sigma,\beta,h)\Bigg|(\mu_1-\mu_0,...,\mu_n-\mu_0)\Bigg)\Bigg)\\
    &=-\frac{1}{2\sigma^2}(\mu_1-\mu_0,...,\mu_n-\mu_0)^{t}\Bigg(\Sigma_{\beta,1}\Sigma_{H,\beta}\Sigma_{\beta,1}^{t}\Bigg)^{-1}(\mu_1-\mu_0,...,\mu_n-\mu_0)\\
    &-\frac{n}{2}\log(2\pi)-n\log(\sigma)-\frac{1}{2}\log(|\Sigma_{\beta,1}\Sigma_{H,\beta}\Sigma_{\beta,1}^{t}|)
\end{split}
\end{equation}
Differentiating with respect to $\sigma$,
\begin{equation}
\begin{split}
    &\frac{\partial\log \Bigg(L\Bigg((\sigma,\beta,h)\Bigg|(\mu_1-\mu_0,...,\mu_n-\mu_0)\Bigg)\Bigg)}{\partial \sigma}\\
   &=\frac{1}{\sigma^3}(\mu_1-\mu_0,...,\mu_n-\mu_0)^{t}\Bigg(\Sigma_{\beta,1}\Sigma_{H,\beta}\Sigma_{\beta,1}^{t}\Bigg)^{-1}(\mu_1-\mu_0,...,\mu_n-\mu_0)-\frac{n}{\sigma}
\end{split}
\end{equation}
We can see that it has a unique local maximum at $(0,\infty)$ and this is given by:
\begin{equation}
    \hat{\sigma}(\beta,H)=\sqrt{\frac{\Bigg(\Sigma_{\beta,1}^{-1}(\mu_1-\mu_0,...,\mu_n-\mu_0)\Bigg)^{t}\Sigma_{H,\beta}^{-1}\Sigma_{\beta,1}^{-1}(\mu_1-\mu_0,...,\mu_n-\mu_0)}{n}}
\end{equation}
Let's consider the profile log-likelihood function of $(\beta,h)$ which is given by:
\begin{equation}
\begin{split}
    \log\Bigg(L_p\Bigg((\beta,H)\Bigg|(\mu_1-\mu_0,...,&\mu_n-\mu_0)\Bigg)\Bigg)=-\frac{n}{2}(1+\log(2\pi))-n\log(\hat{\sigma}(\beta,H))\\
    &-\frac{1}{2}\log(|\Sigma_{\beta,1}\Sigma_{H,\beta}\Sigma_{\beta,1}^{t}|)\\
    &=-\frac{n}{2}(1+\log(2\pi))-\frac{1}{2}\log(|\Sigma_{\beta,\hat{\sigma}(\beta,H)}\Sigma_{H,\beta}\Sigma_{\beta,\hat{\sigma}(\beta,H)}^{t}|)
\end{split}
\end{equation}
When the process is observed at different time scales, we now show the MLEs are scale-invariante in Proposition \ref{Prop2.1}. Considering the amplitude time scale $\Delta$, i.e., $t_i=i\Delta$. We will use the notation $\Sigma_{\beta,1,\Delta}$, $\Sigma_{H,\beta,\Delta}$ and $\hat{\sigma}_{\Delta}(\beta,H)$ to indicate the $\Delta$ scale. The profile log-likelihood is defined as $$f_\Delta(\beta,H):=-n\log(\hat{\sigma}_{\Delta}(\beta,H))-\frac{1}{2}\log(| \Sigma_{\beta,1,\Delta}\Sigma_{H,\beta,\Delta}\Sigma_{\beta,1,\Delta}^t|)$$

 The following results indicate that we can estimate the parameters independently of the scale. The proof is in Appendix A.

\begin{proposition}\label{Prop2.1}
    Let $\Delta_1$ and $\Delta_2$ be two scales. Then,
    
$(i)$ 
    \begin{equation}
       f_{\Delta_2}(\beta,H)=f_{\Delta_1}(\frac{\beta\Delta_2}{\Delta_1},H)
    \end{equation}
\end{proposition}

$(ii)$    Suppose that $(\hat{\beta},\hat{H}) \in (0,\infty)\times(0,1)$ is the MLEs in $\Delta_1$ scale.
Then, $(\frac{\hat{\beta}\Delta_1}{\Delta_2},\hat{H})$ is the MLEs in $\Delta_2$ scale.


Finally, in Appendix B, we present some examples of simulations of finite-dimensional distributions of the position process $\mu_H(t)$ and examine the performance of the MLEs. Empirically, we observe that the maximum likelihood estimators (MLEs) demonstrate consistency. As additional data is added to the likelihood function, the mean square error of the MLEs decreases.

\section{Application to Fin Whale data}\label{practical}

In this section, we apply the animal telemetry model described in Section~\ref{s2} to telemetry data from whales moving over the Gulf of California. Given that the fin whale (\textit{Balaenoptera physalus}) population in the Gulf of California is resident and genetically isolated throughout the year, as reported by the IUCN Marine Mammal Task Force \citep{imma2021gulf}, the movement patterns of individuals, as recorded via telemetry, are expected to exhibit temporal dependence. In such cases, the future position depends on the entire past trajectory.


We assume that the velocity of the trajectories follows a fOU process. In \cite{Jimenez}, telemetry data from eight fin whales moving over the Gulf of California is available. These data were collected between March and September 2001, and include only position and time data. Therefore, we perform parameter inference as described in Section \ref{sec:3.2}, using the maximum likelihood method in finite-dimensional distributions. With respect to the whale identified with the ID $\#1$ we have longitude and latitude data ranging from the dates 03/28/2001 to 07/21/2001 we have a total of 87 records and 59 different days recorded. We consider grouping the data by daily records and on days where we have more than one record we consider their gravest as points of longitude and latitude.


Since we only have longitude and latitude data, we will estimate the parameters via maximum likelihood as described in Section \ref{sec:3.2}. We consider the scale $\Delta=1$. Figure~\ref{LP1} (Appendix C) shows the graph of  $\log(L_p((\beta,H)|(\mu_1-\mu_0,...,\mu_n-\mu_0)) )$ for the latitude and longitude data for the position process. We can see that the likelihood surface exhibits a single maximum. 
We obtain the estimators $(\hat{\sigma}_1 = 1.8060, \hat{\beta}_1 = 6.6453, \hat{H}_1 = 0.3968)$ for the longitude data, and $(\hat{\sigma}_2 = 0.3793, \hat{\beta}_2 = 0.9291, \hat{H}_2 = 0.4431)$ for the latitude data. The Akaike Information Criterion (AIC) is $32.5218$ for the longitude and $31.9367$ for the latitude.

Regarding the whale identified with the ID $\#3$ we have longitude and latitude data ranging from the dates 03/31/2001 to 09/05/2001 we have a total of 217 records and 125 different days recorded. We consider grouping the data by daily records and on days where we have more than one record we consider their gravest as points of longitude and latitude.

  We consider the scale $\Delta=1$. Figure~\ref{LP2} (Appendix C) shows the graph of $\log(L_p((\beta,H)|(\mu_1-\mu_0,...,\mu_n-\mu_0)) )$ for the latitude and longitude data for the position process. We obtain the estimators $(\hat{\sigma_2}=3.0218,\hat{\beta_2} =11.6463,\hat{H_2}=0.74713)$ for latitude and for longitude we selected a value $\beta$ where the profile likelihood function of $\beta$ showed no significant change, since it was not numerically feasible to find the MLE of $\beta$, because its profile likelihood function is flat and increasing. This behavior is consistent with the issue discussed in Remark~\ref{Remarkbeta}. Initially, we set $\beta_{\text{max}}:=400$, the maximum value that $\beta$ can assume, and then select the infimum  $\hat{\beta}\in (0,\beta_{\text{max}}]$ such that for any $\beta \in (\hat{ \beta}, \beta_{\text{max}}]$, the difference evaluating the profile log-likelihood in $\beta$ and $\hat{\beta}$ is not significant ( $<10^{-2}$). We obtained the following estimators of the parameters $(\hat{\sigma_1}=7.2025,\hat{\beta_1}=34.4694,\hat{H_1}=0.5628)$.

In Figure~\ref{fig:lego4} (Appendix C) corresponds to the graph of (a) profile Likelihood $\beta$ and (b) MLE of $\hat{\sigma}(\beta,\hat{H})$. In (b) we can see that if we fix $H$, the MLE of $\sigma$ is increasing. The AIC for this case is $-13.0064$ for the longitude data and $19.0243$ for the latitude data.

With respect to the whale identified with the ID $\#7$, we have longitude and latitude data ranging from the dates 03/27/2001 to 08/26/2001. In total, there are 39 records covering 37 different days. We consider grouping the data by daily records, and on days where we have more than one record, we take their gravest as points of longitude and latitude. We consider the scale $\Delta = 1$. Figure~\ref{LP3} (Appendix~C) shows the graph of $\log(L_p((\beta,H)|(\mu_1-\mu_0, \ldots, \mu_n-\mu_0)))$ for the latitude and longitude data for the position process. We can see that the likelihood surface exhibits a single maximum.

We obtain the estimators $(\hat{\sigma}_1 = 3.3405,\ \hat{\beta}_1 = 1.8032,\ \hat{H}_1 = 0.0150)$ for the longitude data, and $(\hat{\sigma}_2 = 1.0625,\ \hat{\beta}_2 = 0.3488,\ \hat{H}_2 = 0.0309)$ for the latitude data. The AIC is $55.3975$ for the longitude and $59.4762$ for the latitude.

In Table~\ref{T1}, we compare the performance of the model $\mu_H(t)$ with respect to the AIC criterion against the model $\mu_{1/2}(t)$ and the fractional Brownian motion $\sigma W_H(t)$, which is a long-memory process that has been used in various applications, such as modeling dispersion in fluid turbulence (see, e.g., the work of \cite{Lilly}, where fractional Brownian motion is considered and the Matérn process is proposed as an improvement).

We can see that in half of the cases, the model $\mu_H(t)$ performs better under the AIC criterion than the model $\mu_{1/2}(t)$, and the same holds when compared to the model $\sigma W_H(t)$. The analysis of the latitude data for Fin Whale ID~$\#3$ indicates a significant parameter $H > 0.5$ under the AIC criterion when compared to the model $\mu_{1/2}(t)$. This implies that the whale's latitude movement displays transient patterns. Additionally, for the data from Fin Whale~\#7, the model $\mu_H(t)$ outperforms both alternatives under the AIC criterion, in both longitude and latitude. Therefore, we have shown a series of cases in which considering memory structures to model the velocity of animal telemetry data yields better results than the Ornstein–Uhlenbeck velocity model proposed by \cite{John}, and also outperforms the classical long-memory model $\sigma W_H(t)$ under the same AIC criterion.

\begin{table}[htbp]
\centering
\footnotesize
\caption{Comparing the performance of Models $\mu_{H}(t)$, $\mu_{1/2}(t)$, and $\sigma W_H(t)$ applied to the telemetry data of the three Fin Whale trajectories based on the Akaike Information Criterion (AIC).}
\label{T1}
\begin{tabular}{llllllll}
\toprule
Model & AIC & Parameter & MLE & Model & AIC & Parameter & MLE \\
\midrule
\multicolumn{4}{l}{\textbf{Fin Whale \#1: Latitude}} & \multicolumn{4}{l}{\textbf{Fin Whale \#1: Longitude}} \\
\cmidrule(lr){1-8}
$\mu_{H}(t)$ & 32.5218 & $\beta$ & 0.9291 & $\mu_{H}(t)$ & 31.9367 & $\beta$ & 6.6453 \\
& & $\sigma$ & 0.3793 & & & $\sigma$ & 1.806 \\
& & $H$ & 0.4431 & & & $H$ & 0.3968 \\
\cmidrule(lr){1-4} \cmidrule(lr){5-8}
$\mu_{1/2}(t)$ & \textbf{30.6099} & $\beta$ & 1.1169 & $\mu_{1/2}(t)$ & 30.5111 & $\beta$ & 111.1265 \\
& & $\sigma$ & 0.4014 & & & $\sigma$ & 26.4123 \\
\cmidrule(lr){1-4} \cmidrule(lr){5-8}
$\sigma W_H(t)$ & 35.5308 & $H$ & 0.7289 & $\sigma W_H(t)$ & \textbf{30.0202} & $H$ & 0.4436 \\
& & $\sigma$ & 0.2405 & & & $\sigma$ & 0.2435 \\
\midrule
\multicolumn{4}{l}{\textbf{Fin Whale \#3: Latitude}} & \multicolumn{4}{l}{\textbf{Fin Whale \#3: Longitude}} \\
\cmidrule(lr){1-8}
$\mu_{H}(t)$ & 19.0243 & $\beta$ & 11.6463 & $\mu_{H}(t)$ & -13.0064 & $\beta$ & 34.4694 \\
& & $\sigma$ & 3.0218 & & & $\sigma$ & 7.2025 \\
& & $H$ & 0.7471 & & & $H$ & 0.5628 \\
\cmidrule(lr){1-4} \cmidrule(lr){5-8}
$\mu_{1/2}(t)$ & 22.5745 & $\beta$ & 1.62 & $\mu_{1/2}(t)$ & -14.1195 & $\beta$ & 6.4147 \\
& & $\sigma$ & 0.5738 & & & $\sigma$ & 1.4489 \\
\cmidrule(lr){1-4} \cmidrule(lr){5-8}
$\sigma W_H(t)$ & \textbf{17.0925} & $H$ & 0.7618 & $\sigma W_H(t)$ & \textbf{-15.0596} & $H$ & 0.5726 \\
& & $\sigma$ & 0.2574 & & & $\sigma$ & 0.2069 \\
\midrule
\multicolumn{4}{l}{\textbf{Fin Whale \#7: Latitude}} & \multicolumn{4}{l}{\textbf{Fin Whale \#7: Longitude}} \\
\cmidrule(lr){1-8}
$\mu_{H}(t)$ & \textbf{59.4762} & $\beta$ & 0.3488 & $\mu_{H}(t)$ & \textbf{55.3975} & $\beta$ & 1.8032 \\
& & $\sigma$ & 1.0625 & & & $\sigma$ & 3.3405 \\
& & $H$ &  0.0309 & & & $H$ & 0.015 \\
\cmidrule(lr){1-4} \cmidrule(lr){5-8}
$\mu_{1/2}(t)$ & 70.6283 & $\beta$ & 1.3968 & $\mu_{1/2}(t)$ & 70.2808 & $\beta$ & 145.429 \\
& & $\sigma$ & 0.5627 & & & $\sigma$ & 48.9110 \\
\cmidrule(lr){1-4} \cmidrule(lr){5-8}
$\sigma W_H(t)$ & 71.8729 & $H$ & 0.4148 & $\sigma W_H(t)$ & 57.5215 & $H$ & 0.1101 \\
& & $\sigma$ & 0.3828 & & & $\sigma$ & 0.5207 \\
\bottomrule
\end{tabular}
\end{table}

\normalsize

\begin{figure}
\centering
\subfigure[Trajectory Prediction and Velocity Prediction]{\includegraphics[width=140mm,height=12.5cm]{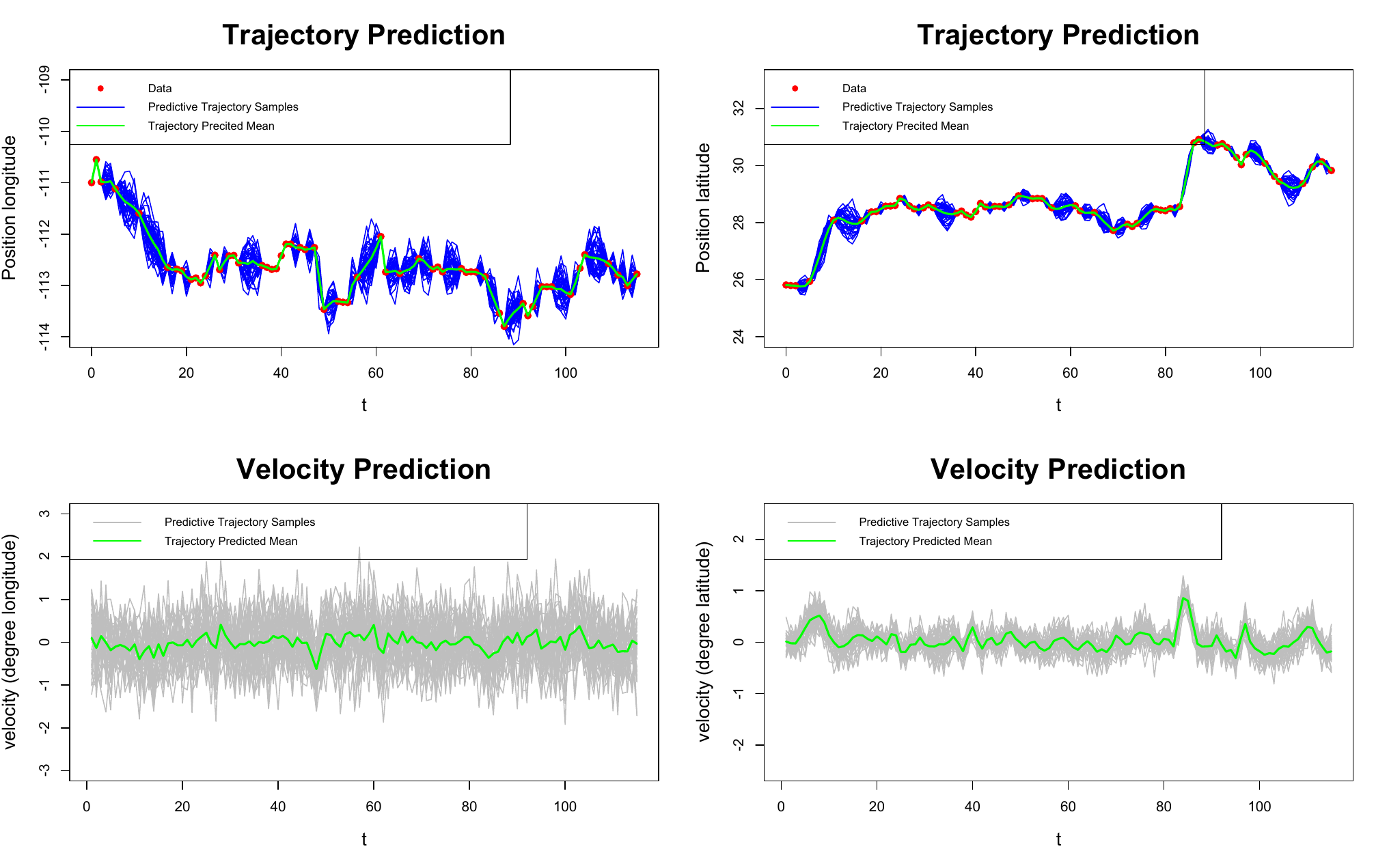}}
\subfigure[Trajectory Prediction]{\includegraphics[width=120mm,height=8cm]{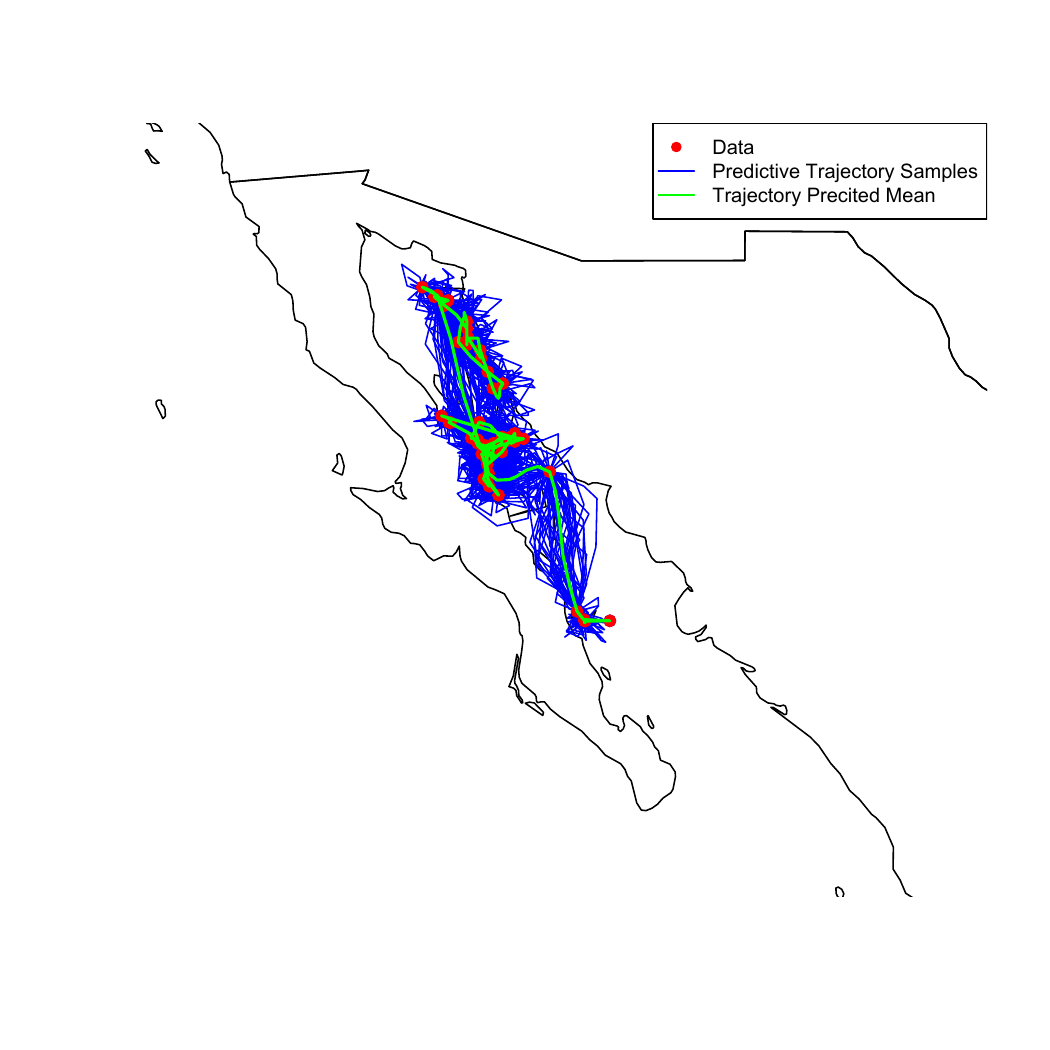}}
\caption{Predicted velocity process for Fin Whale \#1 using positional telemetry data and MLEs. The trajectory predicted mean is based on the mean of 1,000 simulations, providing insights into velocity and position dynamics.} \label{PM1}
\end{figure}

\begin{figure}
\centering
\subfigure[Trajectory Prediction and Velocity Prediction]{\includegraphics[width=140mm,height=12.5cm]{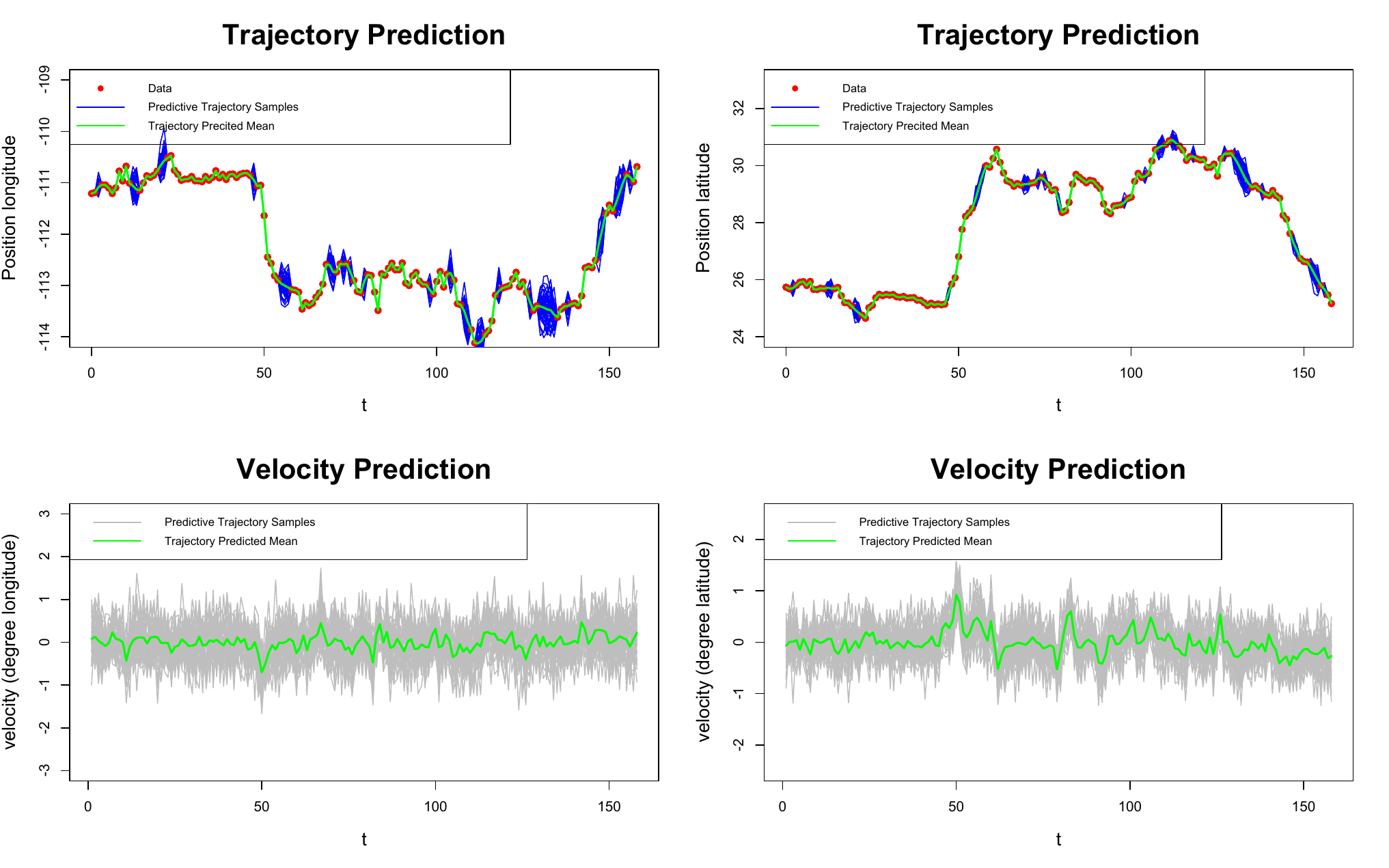}}
\subfigure[Trajectory Prediction]{\includegraphics[width=120mm,height=8cm]{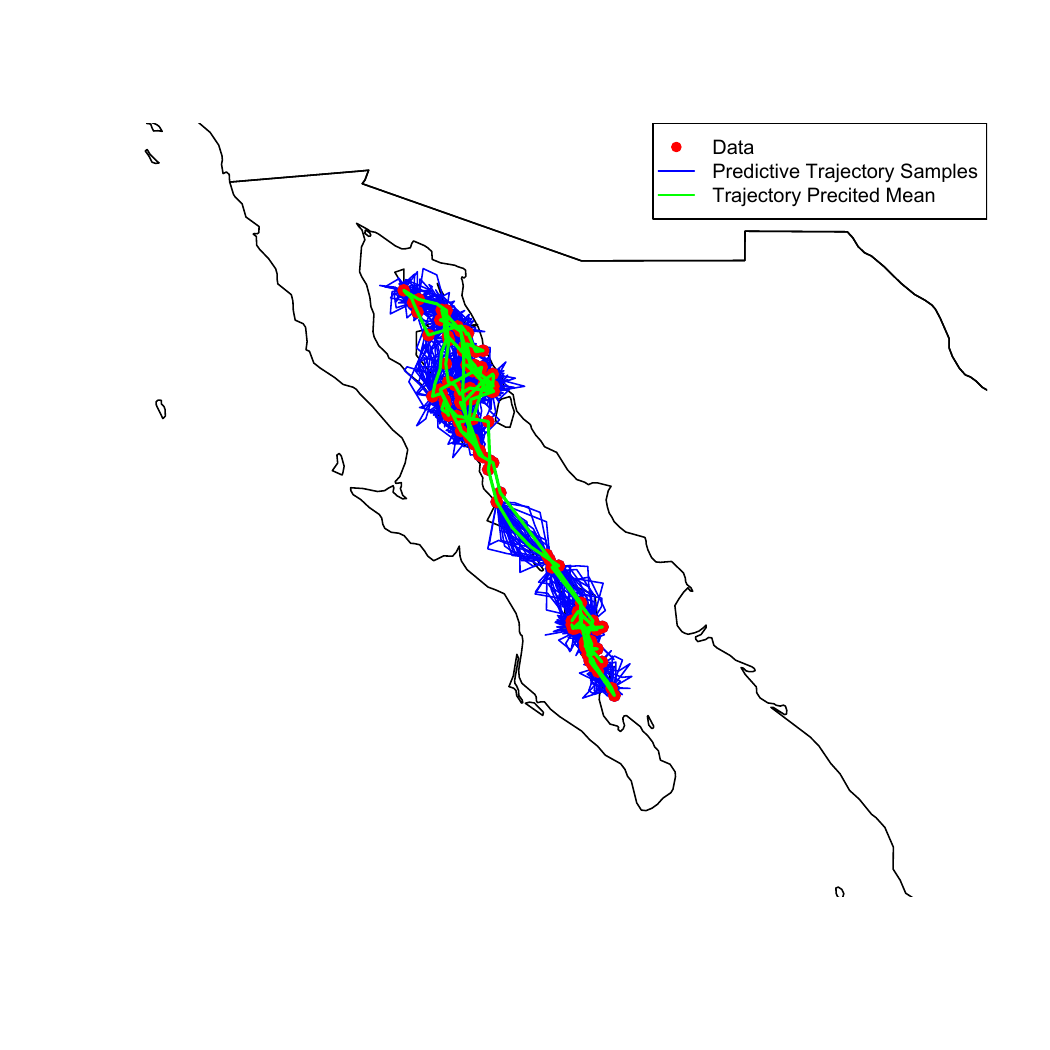}}
\caption{Predicted velocity process for Fin Whale \#3 using positional telemetry data and MLEs. The trajectory predicted mean is based on the mean of 1,000 simulations, providing insights into velocity and position dynamics.}\label{PM2}
\end{figure}

\begin{figure}
\centering
\subfigure[Trajectory Prediction and Velocity Prediction]{\includegraphics[width=140mm,height=12.5cm]{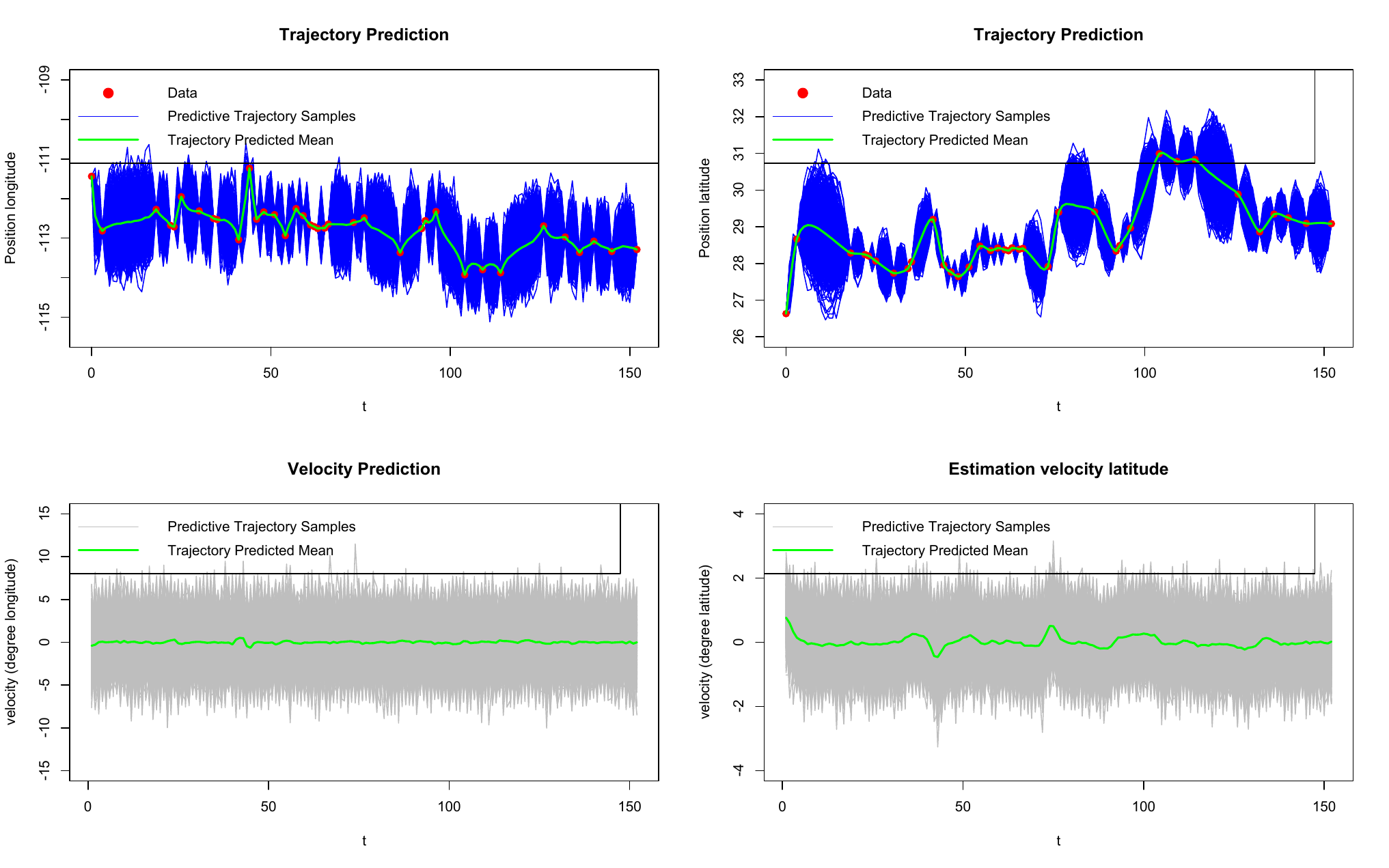}}
\subfigure[Trajectory Prediction]{\includegraphics[width=120mm,height=8cm]{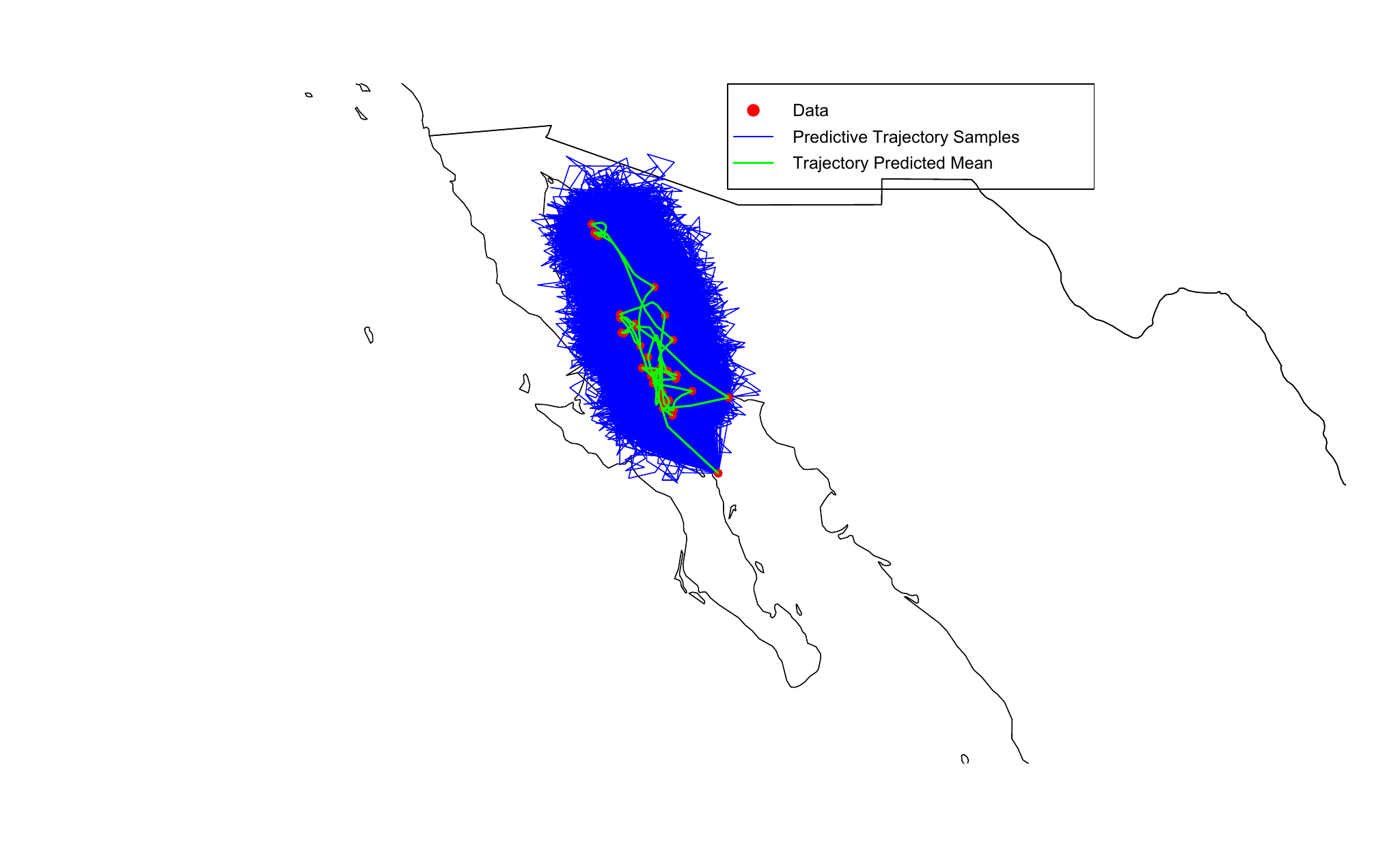}}
\caption{Predicted velocity process for Fin Whale \#7 using positional telemetry data and MLEs. The trajectory predicted mean is based on the mean of 1,000 simulations, providing insights into velocity and position dynamics.}\label{PM3}
\end{figure}

In Figures~\ref{PM1}, \ref{PM2}, and \ref{PM3}, panel~(a) shows simulations of both the velocity and position processes, based on the observed trajectory and the maximum likelihood estimators. The blue lines represent simulated trajectories during periods without observational data.  Panel~(b) displays the reconstructed position process in geographic coordinates (longitude and latitude), providing a joint visualization of the movement on a map.

\newpage

\section{Discussion}\label{sec:discussion}
The process $(\mu_{\frac{1}{2}}(t),v_{\frac{1}{2}}(t))_{t\geq 0}$ is Markovian, which means that future velocity and position are jointly dependent solely on the current state. However, this assumption is often unrealistic for animal trajectories, which typically exhibit local directional preferences. For example, during migration, an animal's movement towards a specific destination reflects a directional trend that is not captured by the conventional Markovian process $(\mu_{\frac{1}{2}}(t),v_{\frac{1}{2}}(t))_{t\geq 0}$. By introducing long-range memory into the model, the future position becomes dependent on the entire trajectory history, thereby providing more accurate predictions for scenarios such as migration.

The real data study presented in Section~\ref{practical} concerning the latitude data of Fin Whale ID $\#3$ reveals a significant parameter $H > 0.5$, as determined by the AIC criterion when compared against the model $\mu_{\frac{1}{2}}(t)$. This suggests that the movement with respect to latitude exhibits transient trends. Moreover, in the case of Fin Whale~\#7, the model $\mu_H(t)$ provided a better fit than both the Ornstein–Uhlenbeck velocity model $\mu_{1/2}(t)$ and the fractional Brownian motion $\sigma W_H(t)$, as measured by the AIC in both latitude and longitude. This supports the relevance of incorporating memory effects in the modeling of velocity, as such models can outperform both classical and long-memory alternatives when evaluated under a model selection criterion like AIC.

In this work, it is assumed that the given movements in longitude and latitude are independent, which is not necessarily true. We have left that case for future work where cross-correlations that respect the structure of the covariance function of each axis can be considered. For example, in \cite{Lavancier, Pierre}, an extension for $d$-dimensional fractional Brownian motion with a correlation structure is discussed. This generalization can be used to introduce correlation structures in the coordinate axes through the equation (\ref{QMH}). The cross-correlation for the position could be calculated directly from the cross-correlation of a $d$-dimensional fractional Brownian motion with a correlation structure. Another interesting generalization would be to consider time-varying coefficients. That is, instead of considering $\beta$ and $\sigma$, consider functions $\beta(t)$ and $\sigma(t)$. 

In addition, it would be of interest to extend the current framework to allow for correlations between individuals. Studying inter-individual dependence through cross-covariance structures could provide insight into group-level movement dynamics, particularly in cases where animals exhibit coordinated or socially influenced behavior.  However, we have left this for future work.

We have observed empirical consistency of the maximum likelihood estimators of the finite-dimensional distributions of $\mu_H(t)$. The theoretical consistency needs to be investigated.

\begin{appendix}

\section{Theoretical proofs}
\textbf{Proof of Proposition \ref{PGH}:}  
First, we will demonstrate that:
    \begin{equation*}
        \mu_{H}(t)=\mu_{H}(0)+v_{H}(0)\Bigg(\frac{1-e^{-\beta t}}{\beta}\Bigg)+\frac{\sigma}{\beta}\int_{0}^{t}(1-e^{\beta(s-t)})dW_{H}(s)
    \end{equation*}
In fact,  in (\cite{fBM1}) they show that the equation: 
\begin{equation}\label{lha2}
    dv_H(t)=-\beta v_{H}(t)dt+\sigma dW_{H}(t), \text{ where } \beta,\sigma>0 \text{ are constants.}
\end{equation}
has a unique trajectory solution, which is given by
\begin{equation}\label{vH1}
    v_H(t)=e^{-\beta t}\Bigg(v_H(0)+\sigma \int_{0}^{t}e^{\beta s}dW_H(s)\Bigg)
\end{equation}
And in Proposition~A.1, they proved that
\begin{equation}\label{lH2}
    \int_{0}^{t}e^{\beta s}dW_H(s)=e^{\beta t}W_{H}(t)-\beta \int_{0}^{t}e^{\beta s}W_{H}(s)ds.
\end{equation}
Since \( (W_H(t))_{t \geq 0} \) is a Gaussian process, equations~(\ref{lha2})--(\ref{vH1}) imply that \( (v_H(t))_{t \geq 0} \) is also Gaussian process, as limits of Gaussian random variables remain Gaussian random variable. Moreover, by equation~(\ref{mpr}) and the same argument, \( \mu_H(t) \) is Gaussian process as well. 

To complete the proof, we compute its covariance function.  
Combining equation~(\ref{vH1}) with Fubini’s Theorem, we obtain

\begin{equation}\label{muH2}
\begin{split}
    \mu_{H}(t)&=\mu_H(0)+\int_{0}^{t}v_{H}(s)\,ds = \mu_{H}(0)+\int_{0}^{t}\left[e^{-\beta s}\left(v_H(0)+\sigma \int_{0}^{s}e^{\beta u}\,dW_H(u)\right)\right]ds\\
    &=\mu_{H}(0)+v_H(0)\left(\frac{1-e^{-\beta t}}{\beta}\right)+\sigma\int_{0}^{t}\int_{0}^{s}e^{\beta (u-s)}\,dW_H(u)\,ds\\
    &=\mu_{H}(0)+v_H(0)\left(\frac{1-e^{-\beta t}}{\beta}\right)+\frac{\sigma}{\beta}\int_{0}^{t}(1-e^{\beta (s-t)})\,dW_H(s).
\end{split}   
\end{equation}

Using equation~(\ref{lH2}), one verifies directly that
\[
\frac{\sigma}{\beta}\int_{0}^{t}(1-e^{\beta (u-t)})\,dW_H(u) = \sigma \int_{0}^{t}e^{\beta(u-t)}W_H(u)\,du.
\]
Substituting into equation~(\ref{muH2}) and computing the covariance leads to
\begin{equation*}
    \begin{split}
        &\operatorname{cov}\left(\frac{\sigma}{\beta}\int_{0}^{t}(1-e^{\beta (u-t)})\,dW_H(u),\,\frac{\sigma}{\beta}\int_{0}^{s}(1-e^{\beta (u-s)})\,dW_H(u)\right)\\
        &=\operatorname{cov}\left(\sigma\int_{0}^{t}e^{\beta (u-t)}W_H(u)\,du,\;\sigma\int_{0}^{s}e^{\beta (u-s)}W_H(u)\,du\right)\\
        &=\sigma^{2}\int_{0}^{t}\int_{0}^{s}e^{\beta(v+u-t-s)}\,\mathbb{E}\left[W_{H}(u)W_{H}(v)\right]\,du\,dv\\
        &=\sigma^{2}\int_{0}^{t}\int_{0}^{s}e^{-\beta v}\,c_{H}(t-v, s-u)\,e^{-\beta u}\,du\,dv,
    \end{split}
\end{equation*}
which completes the proof. \hfill$\square$

Now, we will demonstrate a recursive equations for the velocity and position processes. For the velocity process we have the next results:
\begin{lemma}\label{LT1}
For any $\Delta>0$ we have 
\begin{equation}\label{sh1}
    v_H(t+\Delta)=e^{-\beta \Delta}v_{H}(t)+\sigma\int_{t}^{t+\Delta}e^{\beta(s-(t+\Delta))}dW_H(s).
\end{equation}    
\end{lemma}
\proof 
By the equation (\ref{vH1}) we have that:
\begin{equation}\label{vH}
    \begin{split}
        v_H(t+\Delta)&=e^{-\beta(t+\Delta)}v_H(0)+e^{-\beta(t+\Delta)}\sigma\int_{0}^{t}e^{\beta s}dW_{H}(s)+e^{-\beta(t+\Delta)}\sigma\int_{t}^{t+\Delta}e^{\beta s}dW_{H}(s)\\
        &=e^{-\beta \Delta}v_H(t)+\sigma \int_{t}^{t+\Delta}e^{\beta(s-(t+\Delta))}dW_{H}(s)
    \end{split}
\end{equation}
We conclude the proof. $\hfill \square$

For the position process we have the next results:
\begin{lemma}\label{T2}
    For any $\Delta>0$, we have that

    \begin{equation}\label{sh2}
        \mu_H(t+\Delta)=\mu_H(t)+v_H(t)\Bigg(\frac{1-e^{-\beta \Delta}}{\beta}\Bigg)+\frac{\sigma}{\beta} \int_{t}^{t+\Delta}\Bigg(1-e^{\beta(s-(t+\Delta))}\Bigg)dW_{H}(s)
    \end{equation}
\end{lemma}
\proof 
By equation (\ref{lH2}) we have that 
\begin{equation}\label{muH}
\begin{split}
    \int_{0}^{t}   e^{-\beta u} \int_{0}^{u}e^{\beta s}dW_H(s)du&= \int_{0}^{t}\Bigg[W_{H}(u)-\beta \int_{0}^{u}e^{\beta (s-u)}W_{H}(s)ds\Bigg]du\\
    &=\int_{0}^{t}W_{H}(u)du-\beta \int_{0}^{t}\int_{s}^{t}e^{\beta(s-u)}W_{H}(s)duds\\
    &=\int_{0}^{t}e^{\beta (s-t)}W_{H}(s)ds
   \end{split} 
\end{equation}
The random variable $\int_{t}^{t+\Delta}e^{\beta(s-(t+\Delta))}dW_{H}(s)$ satisfies the equality
\begin{equation}\label{aux1}
    \begin{split}
        \int_{t}^{t+\Delta}&e^{\beta(s-(t+\Delta))}dW_{H}(s)=e^{-\beta(t+\Delta)}\Bigg[\int_{0}^{t+\Delta}e^{\beta s}dW_{H}(s)-\int_{0}^{t}e^{\beta s}dW_{H}(s)\Bigg]\\
        &=e^{-\beta(t+\Delta)}\Bigg[e^{\beta(t+\Delta)}W_{H}(t+\Delta)-\beta\int_{0}^{t+\Delta}e^{\beta s}W_{H}(s)ds-\Bigg(e^{\beta t}W_{H}(t)-\beta\int_{0}^{t}e^{\beta s}W_{H}(s)ds\Bigg)\Bigg]\\
        &=W_{H}(t+\Delta)-e^{-\beta \Delta}W_{H}(t)-e^{-\beta(t+\Delta)}\beta\int_{t}^{t+\Delta}e^{\beta s}W_{H}(s)ds\\
    \end{split}
\end{equation}
For any $\Delta>0$, we have that
\begin{equation}\label{lH4}
\begin{split}
    \int_{0}^{t+\Delta} e^{-\beta s}v_{H}(0)ds-\int_{0}^{t} e^{-\beta s}v_{H}(0)ds&=v_{H}(0)e^{-\beta t}(\frac{1-e^{-\beta \Delta}}{\beta})
    \end{split}
\end{equation}
Therefore, by (\ref{lH2}) and (\ref{aux1}) it is easy to check that
\begin{equation}\label{lH3}
\begin{split}
    &\int_{0}^{t+\Delta}   e^{-\beta u} \int_{0}^{u}e^{\beta s}dW_H(s)du-\int_{0}^{t}   e^{-\beta u} \int_{0}^{u}e^{\beta s}dW_H(s)du\\
   &=\frac{1-e^{-\beta \Delta}}{\beta}e^{-\beta t}\int_{0}^{t}e^{\beta s}dW_{H}(s)+\int_{t}^{t+\Delta}\Bigg(\frac{1-e^{\beta(s-(t+\Delta))}}{\beta}\Bigg)dW_{H}(s)
   \end{split} 
\end{equation}
The proof follows from the equations (\ref{vH1}), (\ref{lH4}) and (\ref{lH3}). $\hfill\square$

\textbf{Proof of Proposition \ref{Prop3.2}}  It follows from the Lemmas \ref{LT1} and \ref{T2}. $\hfill\square$

\textbf{Proof of Proposition \ref{Propv}:} Let's prove equality
\begin{equation}\label{v1a2}
    v_{H}(t_i+\Delta_i)-e^{-\beta \Delta}v_{H}(t_i)=\zeta_{t_i}^{(1)}+\zeta_{t_i}^{(2)}.
\end{equation}
Here,
\begin{equation}\label{z1a2}
  \zeta^{(1)}_{t_i}:=\sigma (W_{H}(t_{i}+\Delta_i)-e^{-\beta \Delta_i}W_{H}(t_i)), 
\end{equation}
and
 \begin{equation}\label{sv1a2}
    \begin{split}
       \zeta^{(2)}_{t_i}&:=-\sigma\beta\int_{t_i}^{t_i+\Delta_i}e^{\beta(s-(t_i+\Delta_i))}W_{H}(s)ds\\
       &=\beta (1-e^{-\beta \Delta_i})\mu_{H}(0)-\beta( \mu_{H}(t_i+\Delta_i)-e^{-\beta \Delta_i}\mu_{H}(t_i))
    \end{split}
\end{equation}
In fact, by the equation (\ref{muH}) we have that:
\begin{equation*}
    \mu_H(t+\Delta)=\mu_H(0)+v_{H}(0)\Bigg(\frac{1-e^{-\beta(t+\Delta)}}{\beta}\Bigg)+\frac{\sigma}{\beta}\int_{0}^{t+\Delta}(1-e^{\beta(s-(t+\Delta))})dW_{H}(s)
\end{equation*}
and
\begin{equation*}
    e^{-\beta \Delta}\mu_{H}(t)=e^{-\beta \Delta}\mu_{H}(0)+v_{H}(0)\Bigg(\frac{e^{-\beta \Delta}-e^{-\beta(t+\Delta)}}{\beta}\Bigg)+\frac{\sigma}{\beta}\int_{0}^{t}\Bigg(e^{-\beta \Delta}-e^{\beta(s-(t+\Delta))}\Bigg)dW_{H}(s)
\end{equation*}
Therefore,
\begin{equation}\label{muHr}
    \begin{split}
        \mu_{H}(t+\Delta)-e^{-\beta \Delta}\mu_{H}(t)&=(1-e^{-\beta \Delta})\mu_{H}(0)+v_{H}(0)\Bigg(\frac{1-e^{-\beta \Delta}}{\beta}\Bigg)\\
        &+\frac{\sigma}{\beta}(W_{H}(t+\Delta)-e^{-\beta \Delta} W_{H}(t))-\frac{\sigma}{\beta}\int_{t}^{t+\Delta}e^{\beta(s-(t+\Delta))}dW_{H}(s)
    \end{split}
\end{equation}
Furthermore, by the equation (\ref{vH})
\begin{equation*}
    \frac{1}{\beta}\Bigg(v_{H}(t+\Delta)-e^{-\beta \Delta}v_{H}(t)\Bigg)=\frac{\sigma}{\beta}\int_{t}^{t+\Delta}e^{\beta(s-(t+\Delta))}dW_{H}(s)
\end{equation*}
Under the assumption $v_H(0)=0$.
\begin{equation}\label{recur}
    \begin{split}
        v_{H}(t+\Delta)-e^{-\beta \Delta}v_H(t)&=\beta (1-e^{-\beta \Delta})\mu_{H}(0)+\sigma (W_H(t+\Delta)-e^{-\beta\Delta}W_{H}(t))\\
        &-\beta (\mu_{H}(t+\Delta)-e^{-\beta \Delta}\mu_{H}(t))\\
    \end{split}
\end{equation}
Finally, The second part of equation (\ref{sv1a2}) follows from equations (\ref{aux1}) and (\ref{muHr}). $\hfill\square$

\textbf{Proof of Proposition~\ref{cov_beta}.} Since \( \mu_H(t) \) is a Gaussian process, it is sufficient to show that \( \lim_{\beta \to \infty} m(t) = \mu_H(0) \) and \( \lim_{\beta \to \infty} Q_{H,\beta,\sigma^{2}}(s,t) = 0 \) for all \( s,t \geq 0 \). Indeed, the convergence \( \lim_{\beta \to \infty} m(t) = \mu_H(0) \) is immediate, and the convergence \( \lim_{\beta \to \infty} Q_{H,\beta,\sigma^{2}}(s,t) = 0 \) follows directly from the Dominated Convergence Theorem.  $\hfill\square$

\textbf{Proof of Proposition \ref{Propy}:} First, we will prove by mathematical induction that 
\begin{equation}\label{z3}
    \mu_i=\mu_0+\sigma \sum_{j=0}^{i-1}y_{j}e^{-\beta (t_i-t_{j+1})}
\end{equation}
In fact, by equation (\ref{sv1a2}) we have that:
\begin{equation}\label{z2}
    -\beta \sigma y_{i}=\beta (1-e^{-\beta \Delta_i})\mu_0-\beta(\mu_{i+1}-e^{-\beta \Delta_i}\mu_i)\phantom{ass}i=0,...,n-1
\end{equation}
In (\ref{z2}) put $i=0$:
\begin{equation*}
    -\beta \sigma y_0=\beta (1-e^{-\beta \Delta_0})\mu_0-\beta (\mu_1-e^{\beta \Delta_0}\mu_0)=\beta\mu_0-\beta \mu_1
\end{equation*}
Then, $\mu_1=\mu_0+\sigma y_0$. We have the equation (\ref{z3}) with $i=1$. We suppose that we have equation (\ref{z3}) for  some $m\in \mathbb{N}$. We proof that we have equation (\ref{z3}) for $m+1$. In fact,
\begin{equation*}
    \begin{split}
        \mu_{m+1}&=(1-e^{-\beta \Delta_m})\mu_0+e^{-\beta\Delta_m} \mu_{m}+\sigma y_{m}\\
        &=(1-e^{-\beta \Delta_m})\mu_0+e^{-\beta\Delta_m}(\mu_0+\sigma\sum_{j=0}^{m-1}y_{j}e^{-\beta (t_{m}-t_{j+1})}) +\sigma y_{m}\\
        &=\mu_0+\sigma \sum_{j=0}^{m-1}y_je^{-\beta(t_{m+1}-t_{j+1})}+\sigma y_{m}=\mu_0+\sigma \sum_{j=0}^{m}y_je^{-\beta(t_{m+1}-t_{j+1})}
    \end{split}
\end{equation*}
Now, it is evident that
\begin{equation*}
\begin{split}
\text{cov}(y_j,y_i)=
 \int_{t_j}^{t_j+\Delta_j}\int_{t_i}^{t_i+\Delta_i}&e^{\beta(u-(t_i+\Delta_i))}e^{\beta(v-(t_j+\Delta_j))}c_{H}(u,v)dudv\\
 &:=\Sigma_{H,\beta}(i+1,j+1), \text{ for }i,j=0,...,n-1 
 \end{split}
\end{equation*}
From where we conclude the proof. $\hfill\square$

\textbf{Proof of Proposition \ref{Prop2.1}:}
(i) Obviously $\Sigma_{\beta,1,\Delta_2}=\Sigma_{\frac{\beta\Delta_2}{\Delta_1},1,\Delta_1}$. Furthermore, by changing the variable $(u^{'},v^{'})=\frac{\Delta_1}{\Delta_2}(u,v)$ we have that
\begin{equation*}
\begin{split}
    \int_{0}^{\Delta_2}\int_{0}^{\Delta_2}&e^{-\beta u}c_{H}(\Delta_2 (i+1)-u,\Delta_2 (j+1)-v)e^{-\beta v}dudv\\
    &=(\frac{\Delta_2}{\Delta_1})^{2+2H}\int_{0}^{\Delta_1}\int_{0}^{\Delta_1}e^{-\frac{\beta\Delta_2}{\Delta_1}u}c_{H}(\Delta_1 (i+1)-u,\Delta_1 (j+1)-v)e^{-\frac{\beta\Delta_2}{\Delta_1}v}dudv
\end{split}
\end{equation*}
Then,
$$\Sigma_{H,\beta,\Delta_2}=(\frac{\Delta_2}{\Delta_1})^{2+2H}\Sigma_{H,\frac{\beta \Delta_2}{\Delta_1},\Delta_1}$$
Therefore,  
\begin{equation*}
    \begin{split}
        &\hat{\sigma}_{\Delta_2}(\beta,H)=\sqrt{\frac{(\mu_1-\mu_0,...,\mu_n-\mu_0)^{t}(\Sigma_{\beta,1,\Delta_2}\Sigma_{H,\beta,\Delta_2}\Sigma_{\beta,1,\Delta_2}^{t})^{-1}(\mu_1-\mu_0,...,\mu_n-\mu_0)}{n}}\\
        &=(\frac{\Delta_1}{\Delta_2})^{1+H}\sqrt{\frac{(\mu_1-\mu_0,...,\mu_n-\mu_0)^{t}(\Sigma_{\frac{\beta\Delta_2}{\Delta_1},1,\Delta_1}\Sigma_{H,\frac{\beta\Delta_2}{\Delta_1},\Delta_1}\Sigma_{\frac{\beta\Delta_2}{\Delta_1},1,\Delta_1}^{t})^{-1}(\mu_1-\mu_0,...,\mu_n-\mu_0)}{n}}
    \end{split}
\end{equation*}
Finally,
\begin{equation*}
    \begin{split}
        f_{\Delta_2}(\beta,H)&=-n(1+H)\log(\frac{\Delta_1}{\Delta_2})-n\log(\hat{\sigma}_{\Delta_1}(\frac{\beta \Delta_2}{\Delta_1},H))-n(1+H)\log(\frac{\Delta_2}{\Delta_1})\\
        &-\frac{1}{2}\log(|\Sigma_{\frac{\beta \Delta_2}{\Delta_1},1,\Delta_1}\Sigma_{H,\frac{\beta\Delta_2}{\Delta_1},\Delta_1}\Sigma_{\frac{\beta \Delta_2}{\Delta_1},1,\Delta_1}|)\\
        &=f_{\Delta_1}(\frac{\beta\Delta_2}{\Delta_1},H)
    \end{split}
\end{equation*}
$(ii)$ It is immediate of (i). $\hfill \square$

\section{Simulation study}

\subsection{Simulation}

\begin{figure}[H]
\centering
\subfigure[$(H_1=0.04,H_2=0.02)$]{\includegraphics[width=63mm]{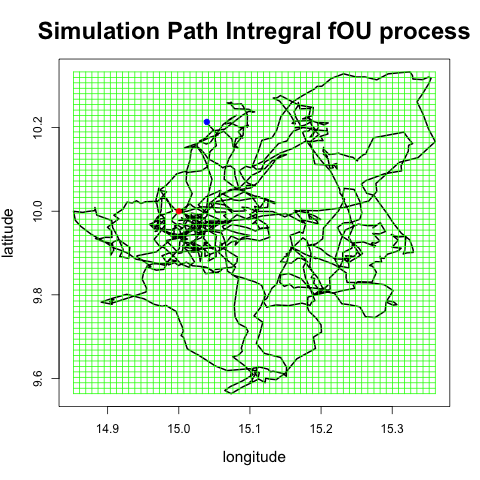}}\
\subfigure[$(H_1=0.22,H_2=0.18)$]{\includegraphics[width=63mm]{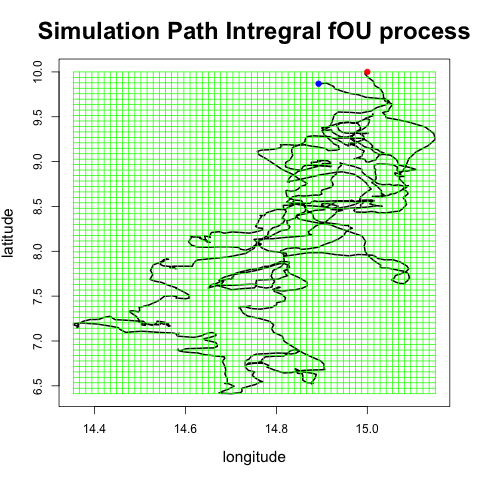}}\vspace{1mm}
\subfigure[$(H_1=0.3,H_2=0.5)$]{\includegraphics[width=63mm]{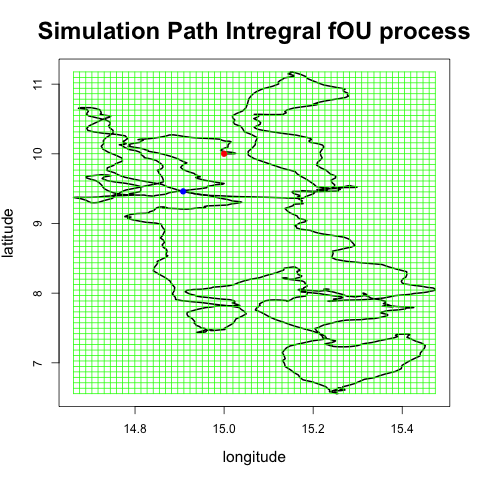}}
\subfigure[$(H_1=0.5,H_2=0.5)$]{\includegraphics[width=63mm]{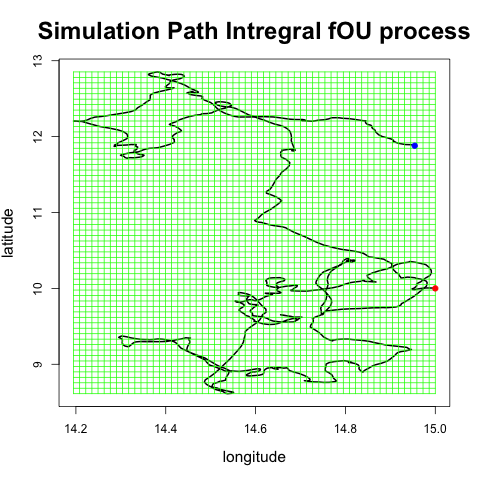}}
\caption{Simulation of two independent integral fOU processes (one on each axis): In the simulations we consider $T=20$ and $\Delta=\frac{1}{50}$ and the parameters are $(\sigma_1=\sqrt{3},\beta_1=8.1,\mu_1(0)=15,\sigma_2=\sqrt{6},\beta_2=2.7,\mu_2(0)=10)$.} \label{F1}
\end{figure}

\begin{figure}[H]
\centering
\subfigure[$(H_1=0.7,H_2=0.5)$]{\includegraphics[width=63mm]{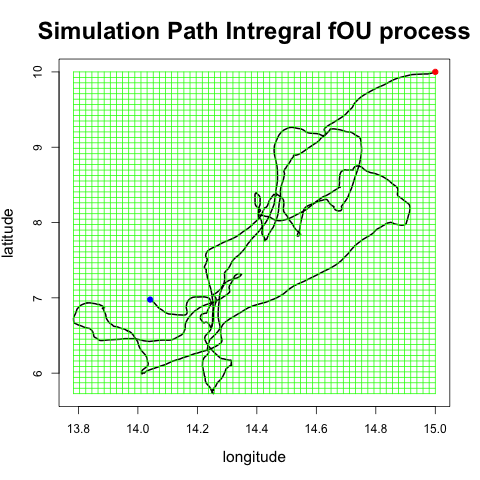}}
\subfigure[$(H_1=0.82,H_2=0.76)$]{\includegraphics[width=63mm]{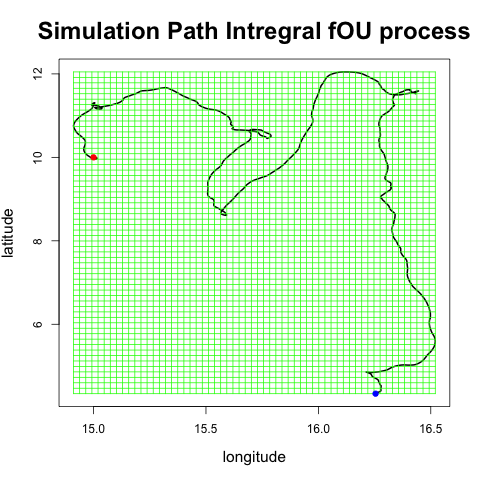}}\vspace{1mm}
\subfigure[$(H_1=0.94,H_2=0.95)$]{\includegraphics[width=63mm]{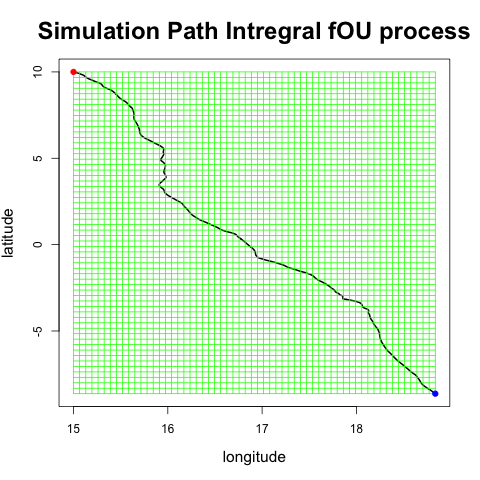}}
\subfigure[$(H_1=0.03,H_2=0.92)$]{\includegraphics[width=63mm]{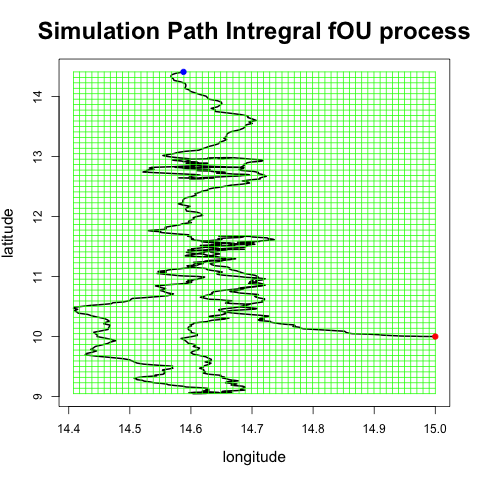}}
\caption{Simulation of two independent integral fOU processes (one on each axis): In the simulations we consider $T=20$ and $\Delta=\frac{1}{50}$ and the parameters are $(\sigma_1=\sqrt{3},\beta_1=8.1,\mu_1(0)=15,\sigma_2=\sqrt{6},\beta_2=2.7,\mu_2(0)=10)$.} \label{F2}
\end{figure}

Figures \ref{F1} and \ref{F2} present simulated trajectories of the positioning process generated using Algorithm \ref{alg:pos}. On each axis, we consider an independent integral fOU process. The trajectories are evaluated at times $0=t_0<t_1<...<t_n=T$ with $T = 20$, $n=1000$, and $\Delta:=t_{i}-t_{i-1}=\frac{1}{50}$ for $i=1,...,n$. The parameters are $(\sigma_1=\sqrt{3},\beta_1=8.1,\mu_1(0)=15,\sigma_2=\sqrt{6},\beta_2=2.7,\mu_2(0)=10)$. Due to the correlation of the increments, we observe recurrent trajectories on the axis where the Hurst parameter $H \in (0, \frac{1}{2})$ and transient trajectories on the axis where the Hurst parameter $H \in (\frac {1}{2},1)$. Consequently, in Figure \ref{F1} (a), (b) and (c) and in Figure \ref{F2} (a), (b) and (c) we can see trajectories with recurrent and transitory trends respectively. Figure \ref{F1} (d) shows a trajectory when $(H_1=H_2=0.5)$. This is the process used to model animal telemetry data in \cite{John}. Finally, Figure \ref{F2} (d) shows a trajectory that has a recurrence component and a transitory component in the axes. We can see recurring trend on the longitude axis and transient trend on the latitude axis.

\subsection{Inference to simulated data}
\begin{figure}[H]
\centering
\subfigure[longitude]{\includegraphics[width=63mm]{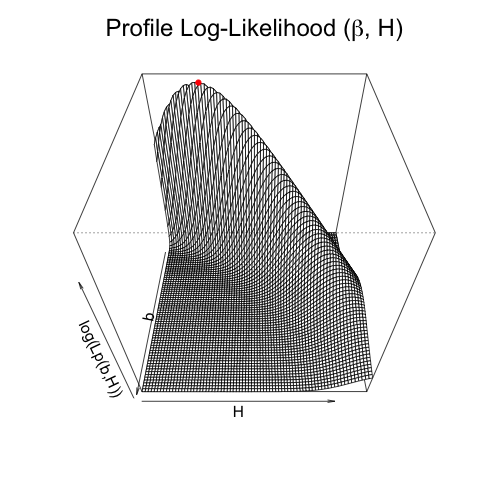}}\hspace{3mm}
\subfigure[latitude]{\includegraphics[width=63mm]{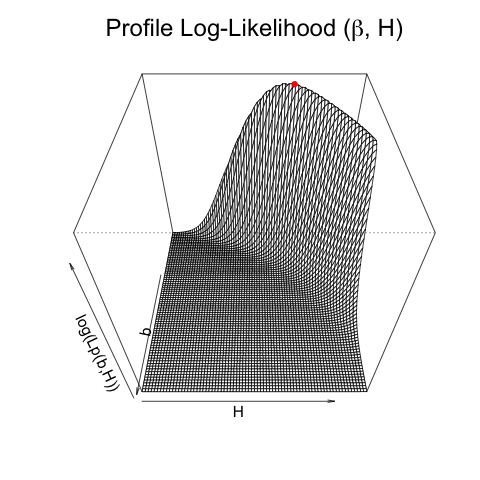}}
\caption{Profile Likelihood $(\beta,H)$ for longitude and latitude (simulated data). The red points indicate the maximum value of the Log-likelihood $(\beta,H)$ function of each axis.} \label{F4}
\end{figure}

\begin{figure}[H]
\centering
\includegraphics[width=14cm,height=8cm]{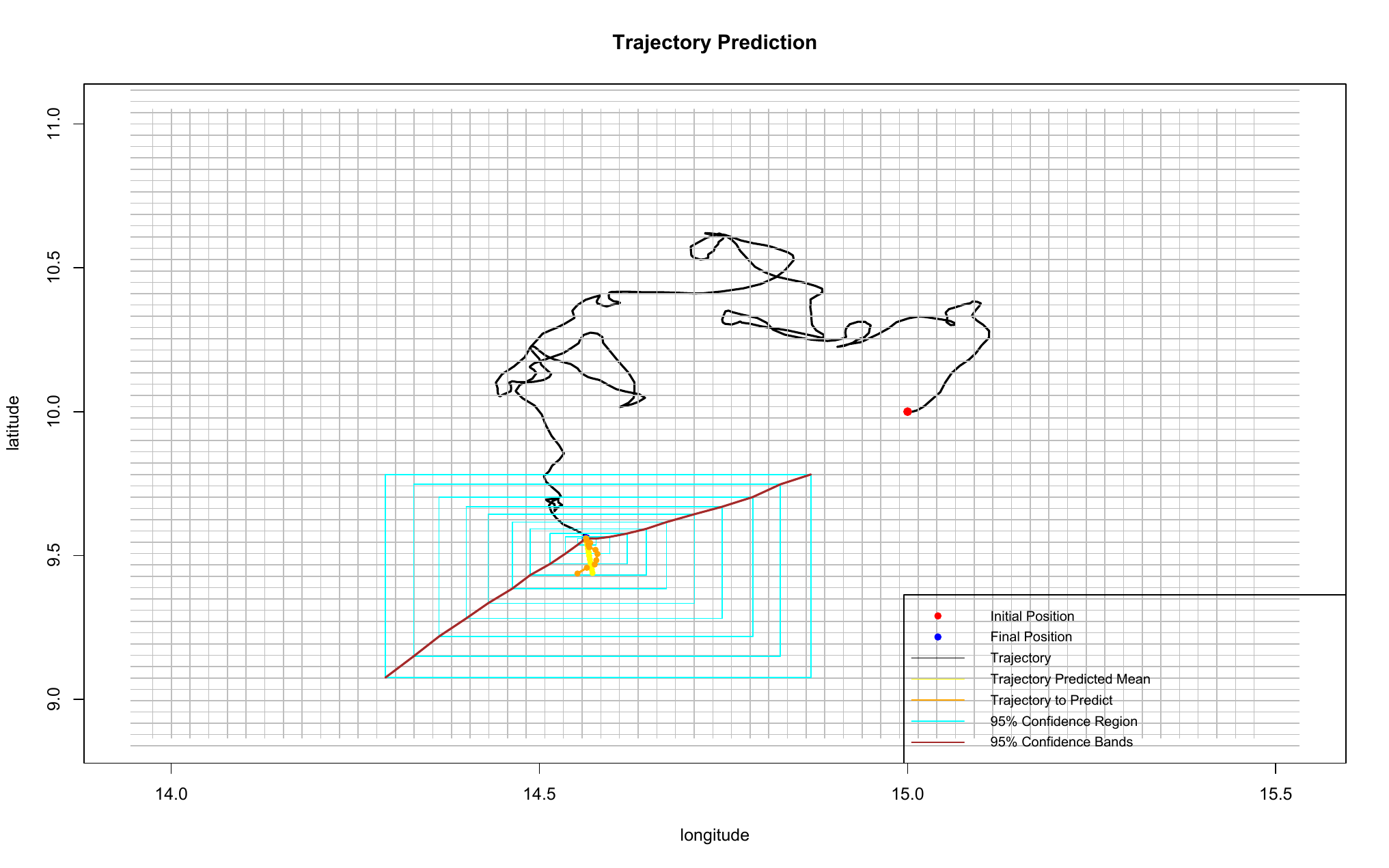}
\caption{Trajectory Predictions for simulated data with the following parameters $T=10$, $n=300$, $\Delta=\frac{1}{30}$, $\mu_1(0)=15$, $\mu_2(0)=10$, $( \sigma_1=\sqrt{3},\beta_1=12,H_1=0.56)$ and $(\sigma_2=\sqrt{7},\beta_2=6,H_2=0.75)$.} \label{FP}
\end{figure}

\begin{figure}[H]
\centering
\includegraphics[width=14cm,height=8cm]{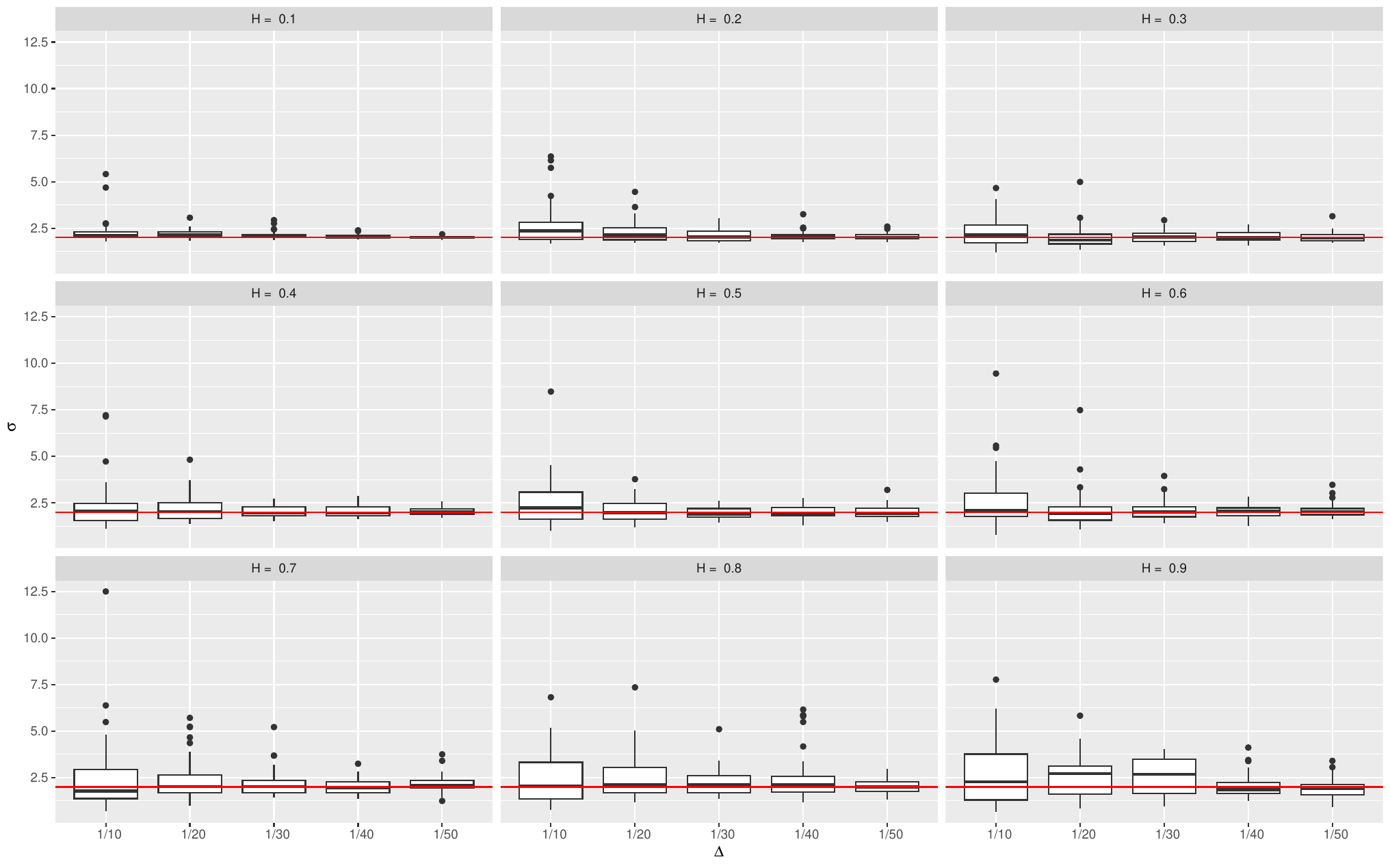}
\caption{Maximum likelihood estimator of $\sigma$ with $\sigma=2$, $\beta=3$, $H=0.1, 0.2,..., 0.9$ and $\Delta=1/10,1/20, ...,1/50$. The red line indicates the value of $\sigma$.} \label{fig:paht}
\end{figure}

\begin{figure}[H]
\centering
\includegraphics[width=14cm,height=8cm]{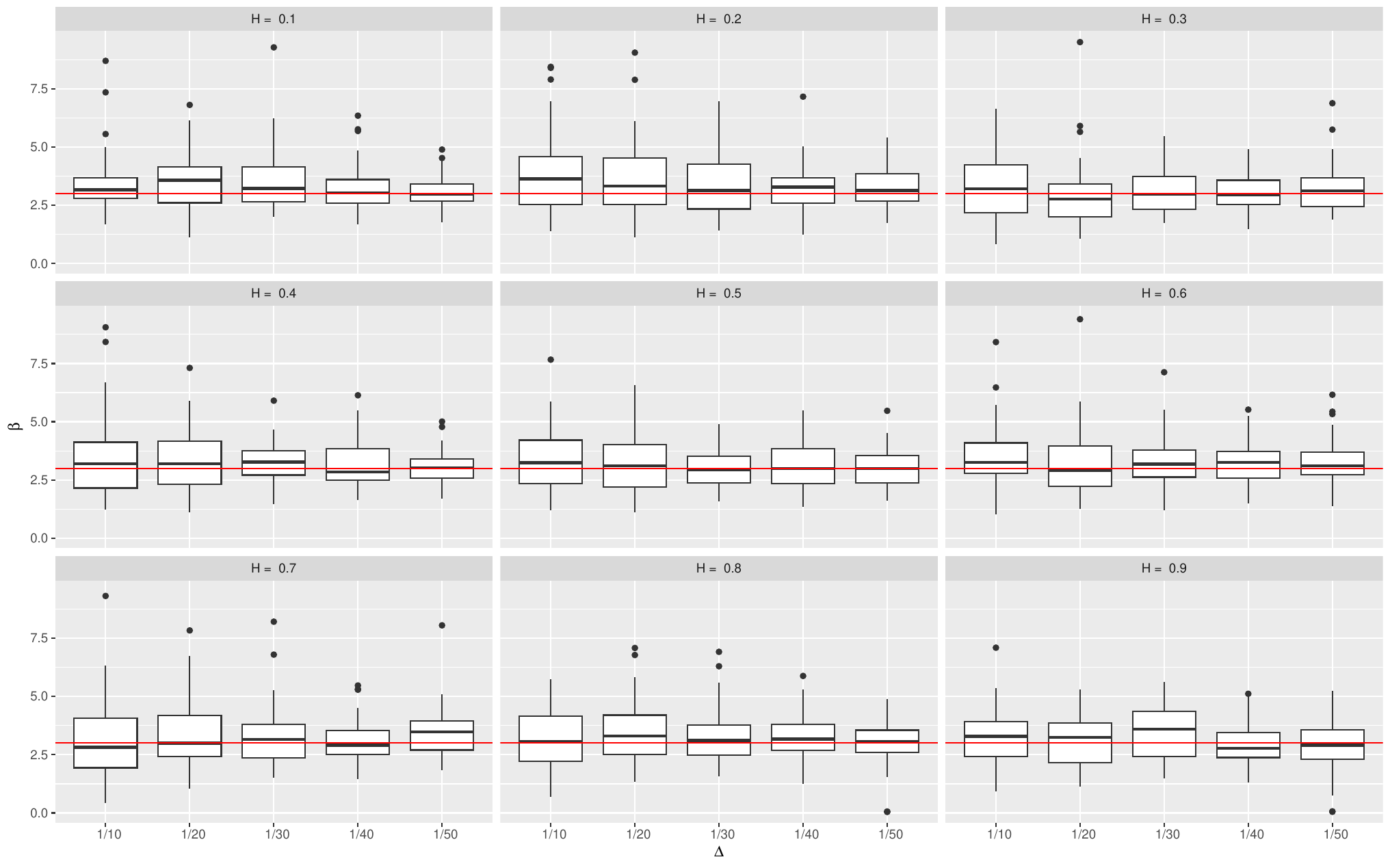}
\caption{Maximum likelihood estimator of $\beta$ with $\sigma=2$, $\beta=3$, $H=0.1, 0.2,..., 0.9$ and $\Delta=1/10,1/20, ...,1/50$. The red line indicates the value of $\beta$.} 
\label{fig:paht2}
\end{figure}

\begin{figure}[H]
\centering
\includegraphics[width=14cm,height=8cm]{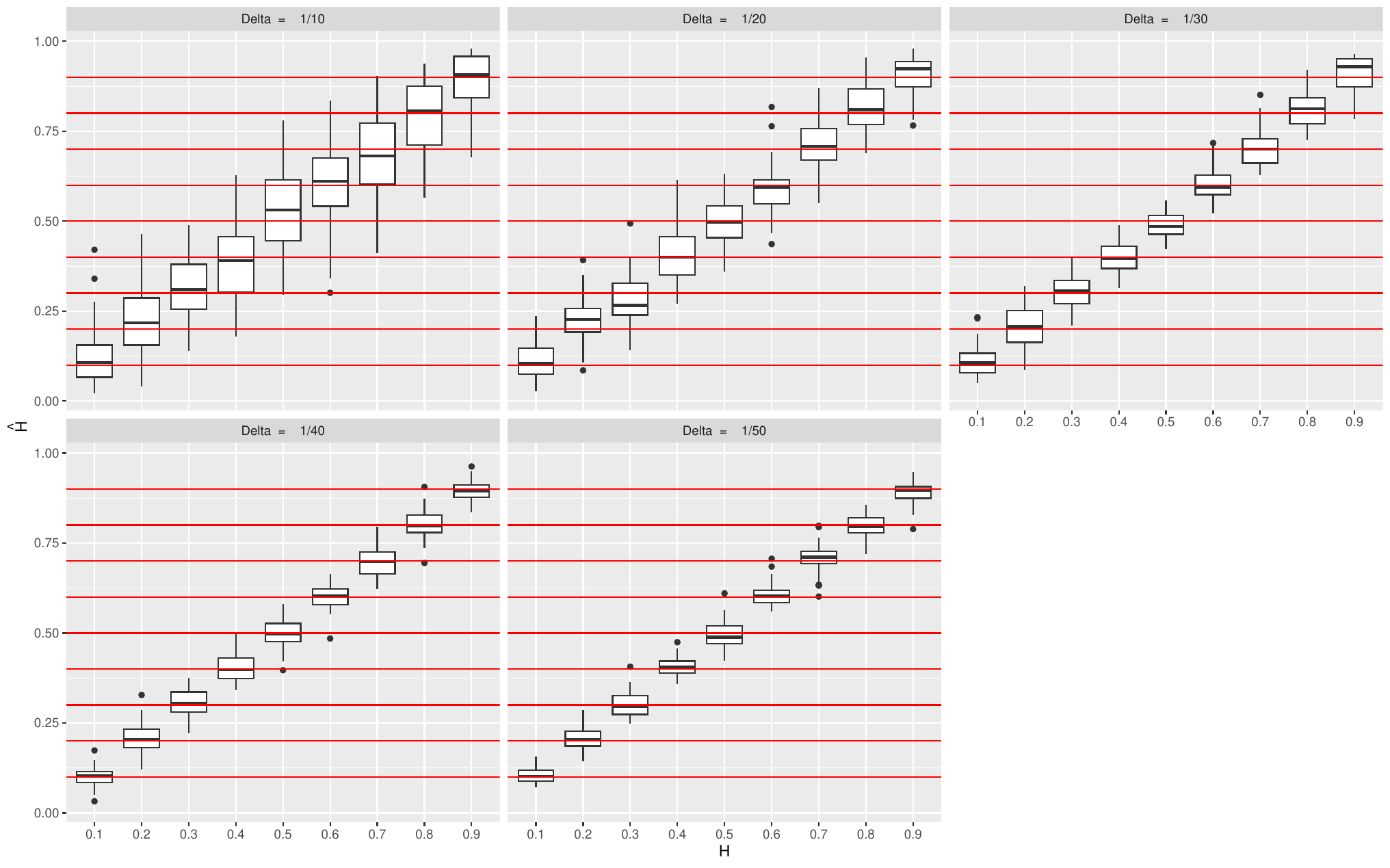}
\caption{Maximum likelihood estimator of $H$ with $\sigma=2$ $\beta=3$ and $\Delta=1/10,1/20, ...,1/50$. The red line indicates the value of $H$. } 
\label{fig:paht3}
\end{figure}

We perform a simulation with the following parameters $T_r=\frac{31}{3}$, $n_r=310$, $\Delta=\frac{1}{30}$, $\mu_1(0)=15$, $\mu_2(0)=10$, $( \sigma_1=\sqrt{3},\beta_1=12,H_1=0.56)$ and $(\sigma_2=\sqrt{7},\beta_2=6,H_2=0.75)$. For parameter inference, we used the first $n=300$ data points, and the remaining 10 data points were used to calculate the mean square error between the trajectory predicted mean and the real data. By Proposition 4.1 we can consider $\Delta=1$ for the estimate. Then, $T=n=300$, we obtain the estimators $(\hat{\sigma_1}=0.00801,\hat{\beta_1}=0.25802,\hat{H_1}=0.47547)$ and $(\hat{\sigma_2}=0.00695,\hat{\beta_2} =0.19274,\hat{H_2}=0.70583)$. In Figure \ref{F4} shows the graph of the profile likelihood function of $(\beta_1,H_1)$ (longitude, panel (a)) and $(\beta _2,H_2)$ (latitude, panel (b)). Considering the original scale $\Delta=\frac{1}{30}$. We have that $T=10$, $n=300$ and by Proposition 4.1 we obtain the estimators $(\hat{\sigma_1}=1.14603,\hat{\beta_1}=7.74068,\hat{H_1}=0.47547)$ and $(\hat{\sigma_2}=2.20597,\hat{\beta_2}=5.78245,\hat{H_2}=0.70583)$. The profile likelihood $(\beta,H)$ function has a single global maximum in both cases.
Using the Algorithm \ref{alg:position}, we can make trajectory predictions in the following 10 steps. Figure \ref{FP} depicts the prediction of the next 10 steps. We conducted 3000 simulations, using the trajectory predicted mean (yellow line) for the prediction. We compared the results using mean square error with the real data (orange line) and obtained a mean square error of 0.00006 for longitude data, and 0.00026 for latitude data. Therefore, we consider that the model is capable of accurately predicting future trajectory values based on simulated data.
Additionally, we perform a simulation analysis to empirically demonstrate the consistency of the maximum likelihood estimators. Here, we consider $\mu_H(0)=10$, $(\sigma=2,\beta=3,H=0.1,0.2,...,0.9)$, $\Delta=1/10,1/ 20,...,1/50$, and $T=10$. We perform inference on 50 simulations of each parameter combination. In Figure \ref{fig:paht}, Figure \ref{fig:paht2}, and Figure \ref{fig:paht3}, we empirically observe the consistency of the maximum likelihood estimators (MLE). As more data is incorporated into the likelihood function, the MLEs exhibit a reduction in mean squared error.

\section{Auxiliary Figures for the Application.}

\begin{figure}[H]
\centering
\subfigure[longitude]{\includegraphics[width=63mm]{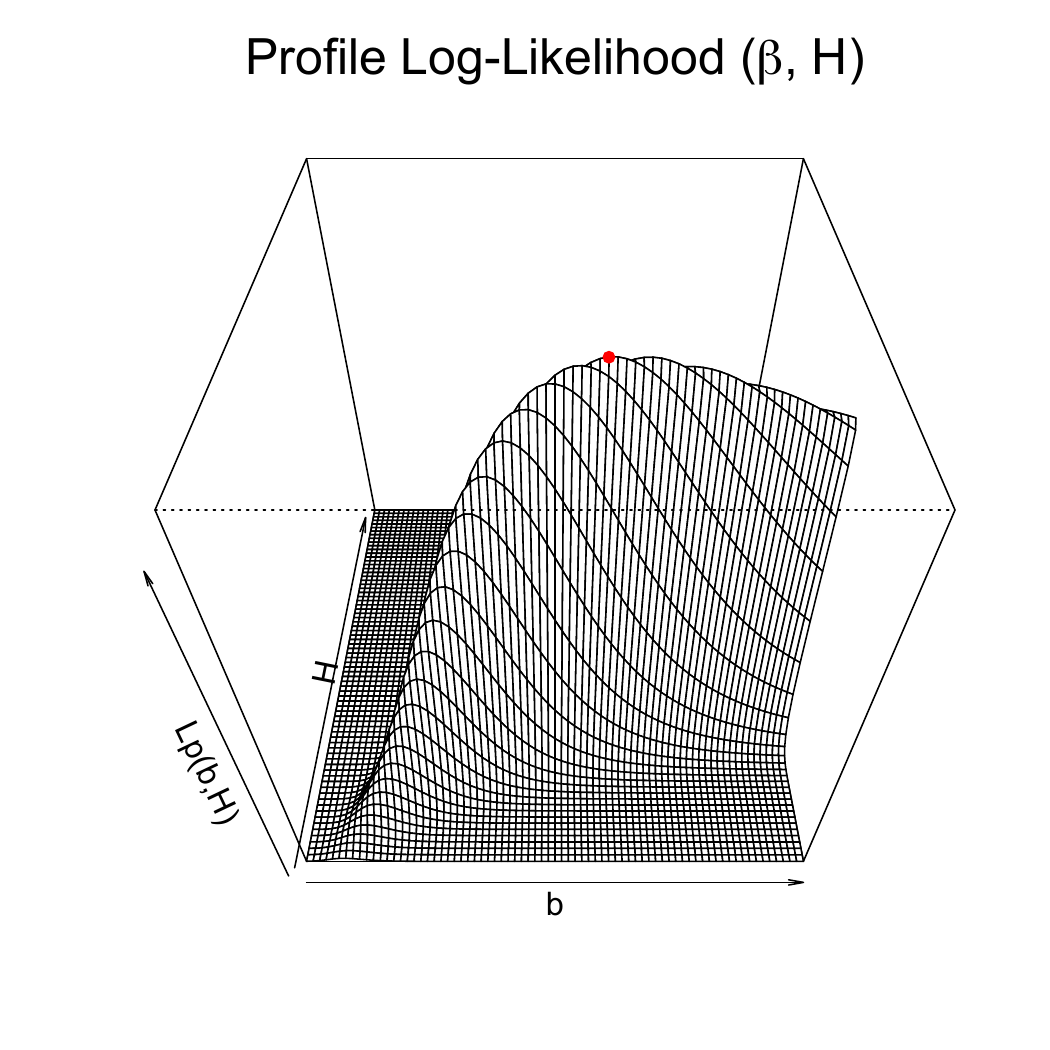}}\hspace{3mm}
\subfigure[latitude]{\includegraphics[width=63mm]{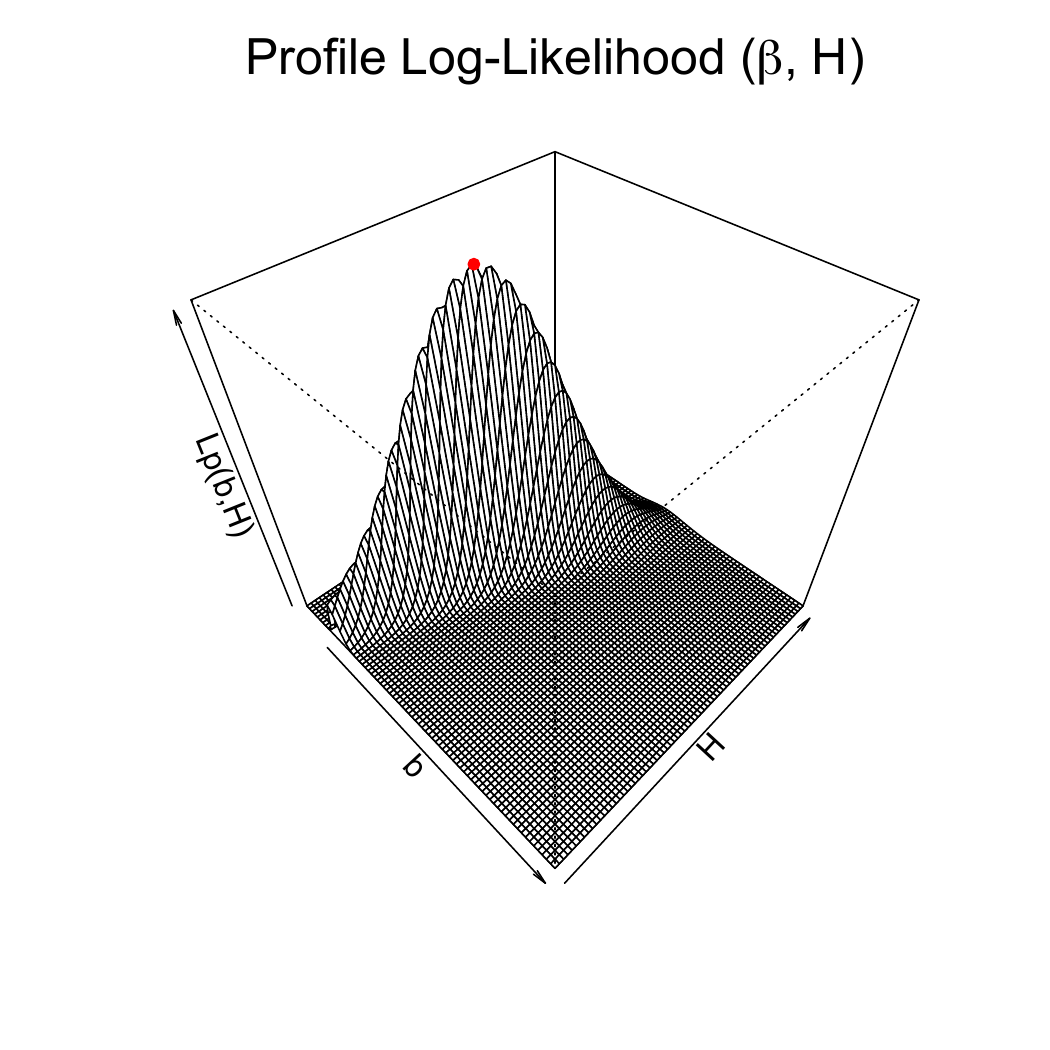}}
\caption{Profile Likelihood $(\beta,h)$ for longitude and latitude (Fin Whale $\#1$). The red points indicate the maximum value of the Log-likelihood $(\beta,H)$ function of each axis.} \label{LP1}
\end{figure}
\begin{figure}[H]
\centering
\subfigure[longitude]{\includegraphics[width=63mm]{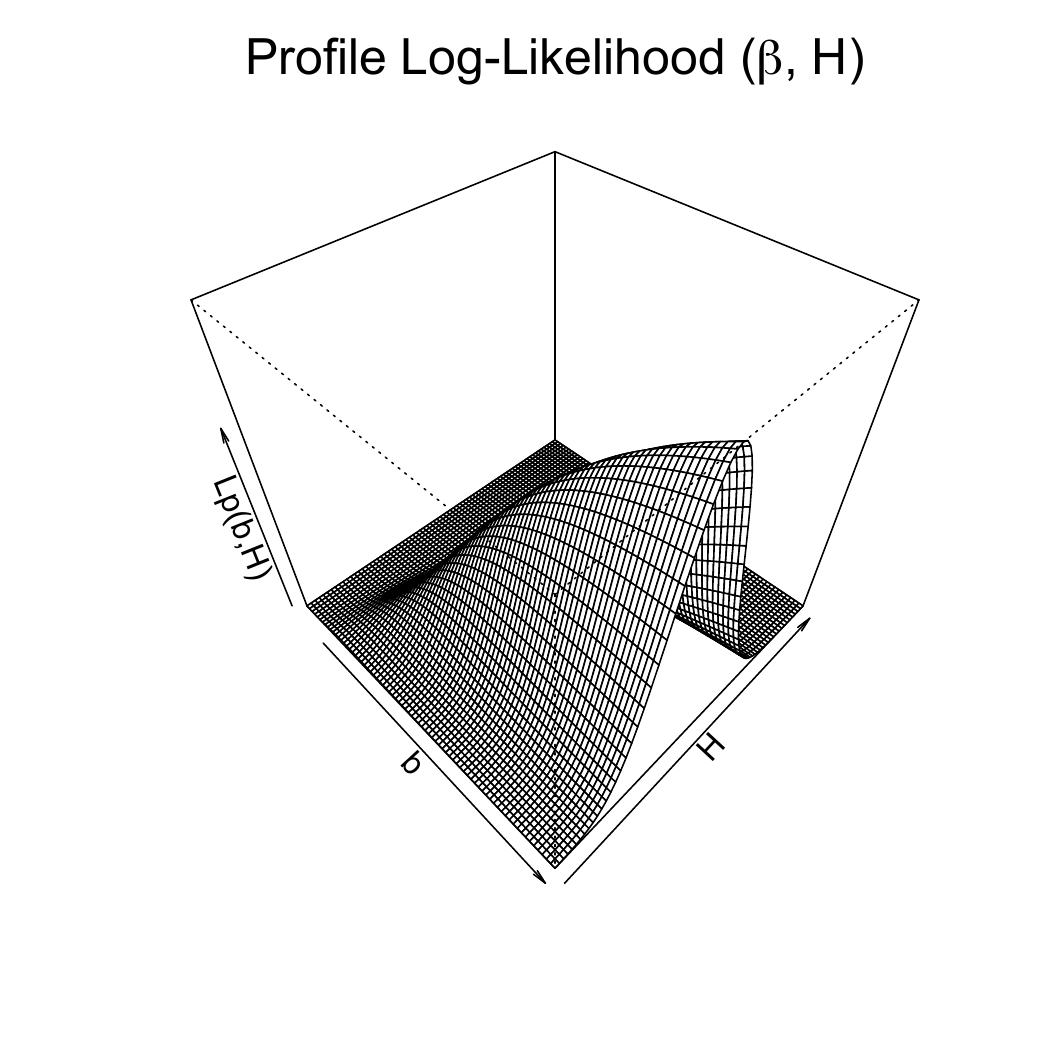}}\hspace{3mm}
\subfigure[latitude]{\includegraphics[width=63mm]{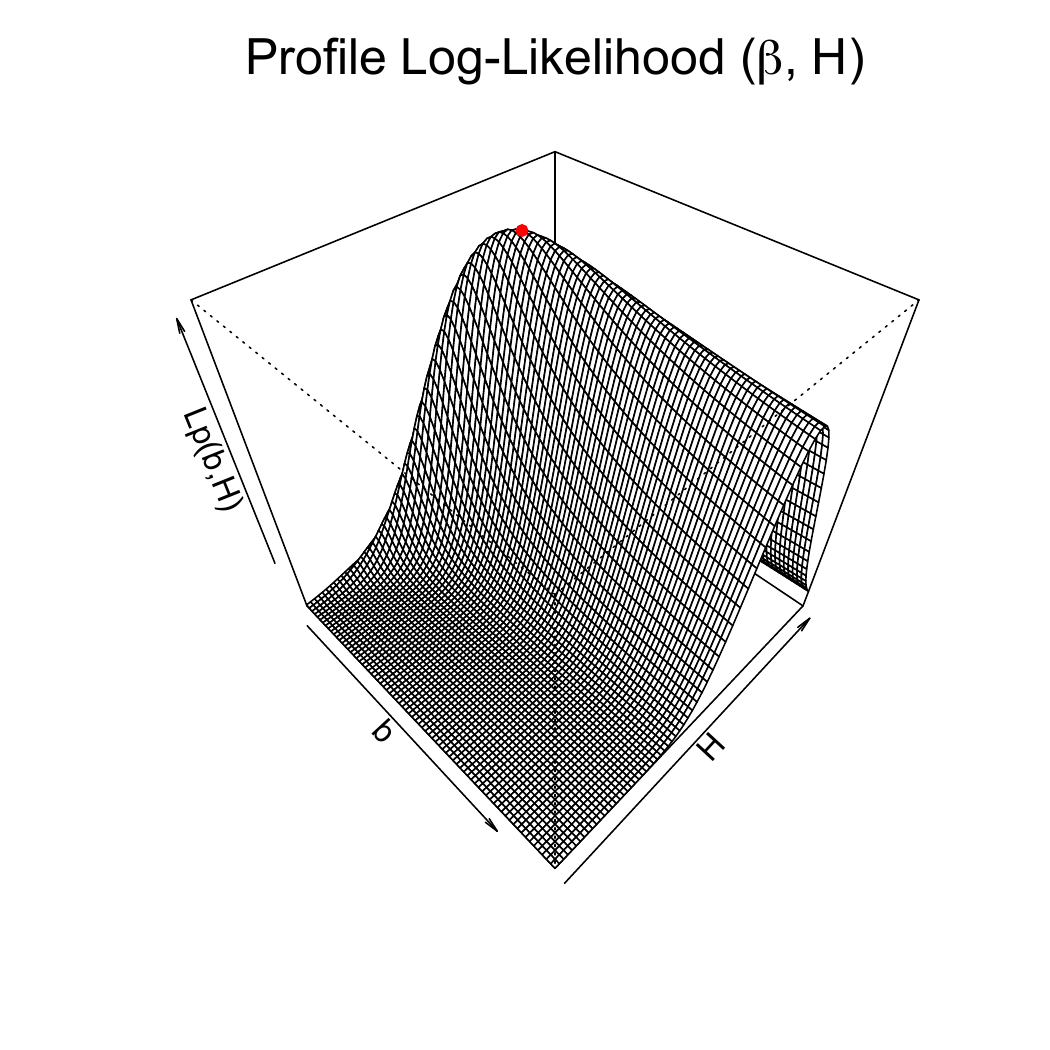}}
\caption{Profile Likelihood $(\beta,h)$ for longitude and latitude (Fin Whale $\#3$). The red points indicate the maximum value of the Log-likelihood $(\beta,H)$ function of each axis. } \label{LP2}
\end{figure}
\begin{figure}[H]
\centering
\subfigure[$L_p(\beta)$]{\includegraphics[width=6.3cm, height=4cm]{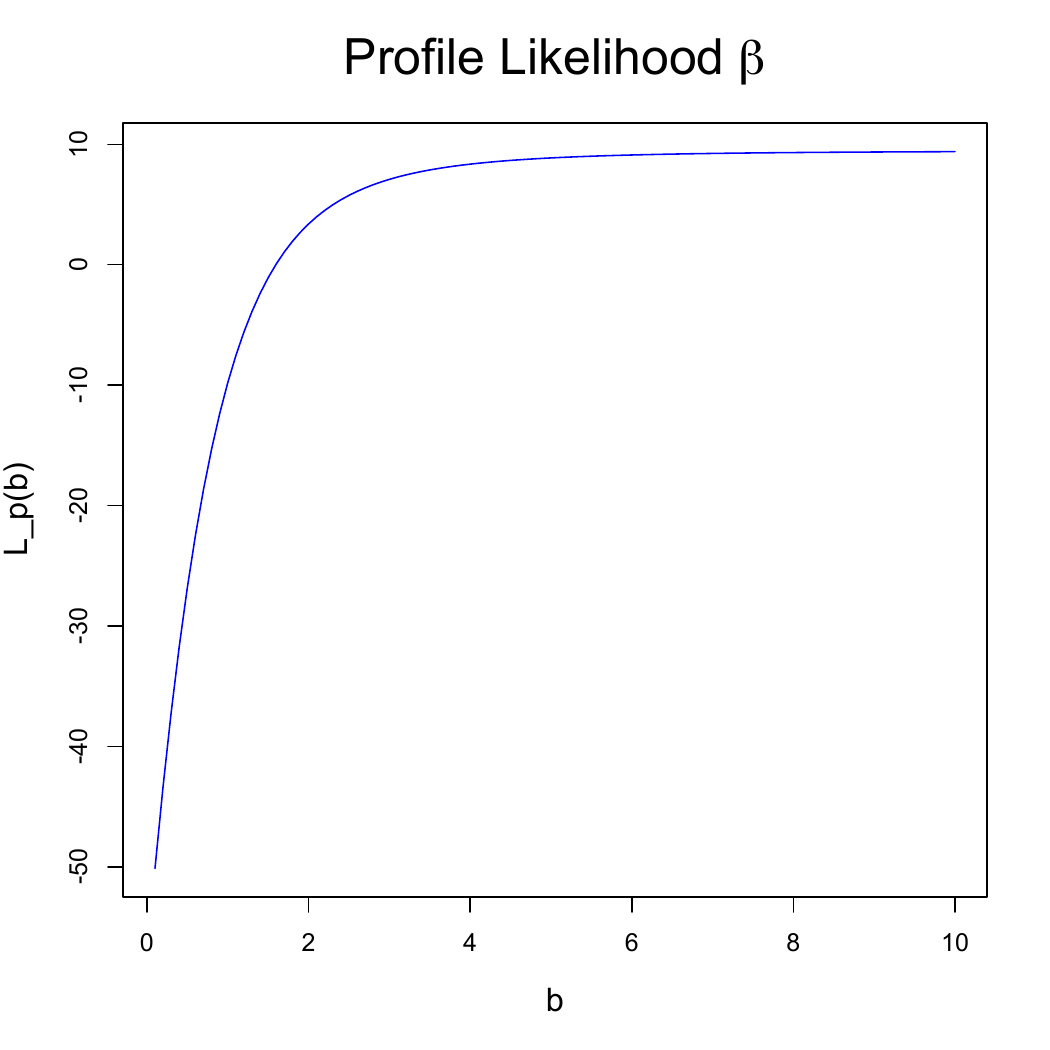}}\hspace{3mm}
\subfigure[$\hat{\sigma}(\beta,\hat{H})$]{\includegraphics[width=6.3cm, height=4cm]{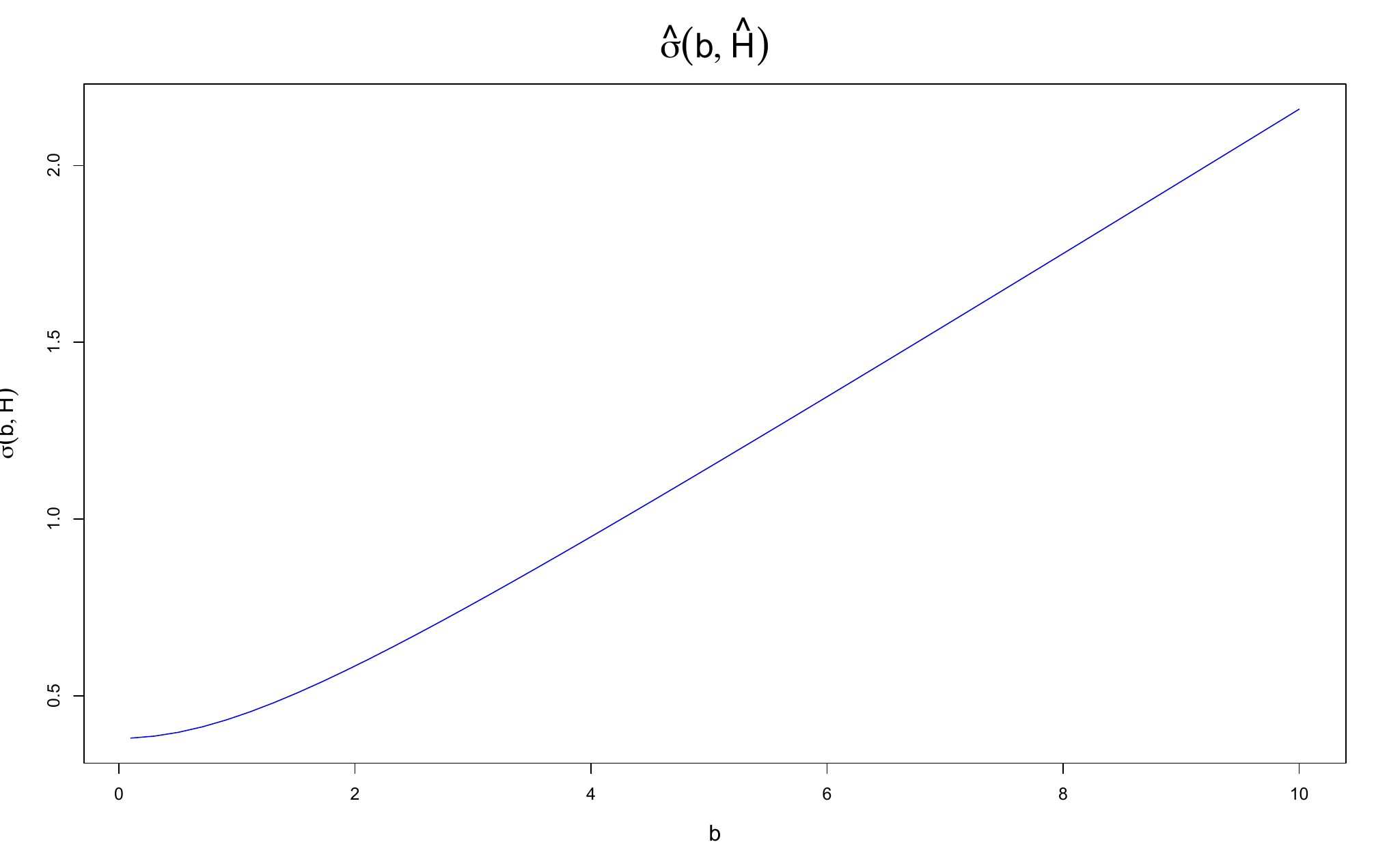}}\hspace{3mm}
\caption{(a) Profile Likelihood $\beta$ for Fin Whale $\#3$ and (b) MLE of $\hat{\sigma}(\beta,\hat{H})$.} \label{fig:lego4}
\end{figure}
\begin{figure}[H]
\centering
\subfigure[longitude]{\includegraphics[width=63mm]{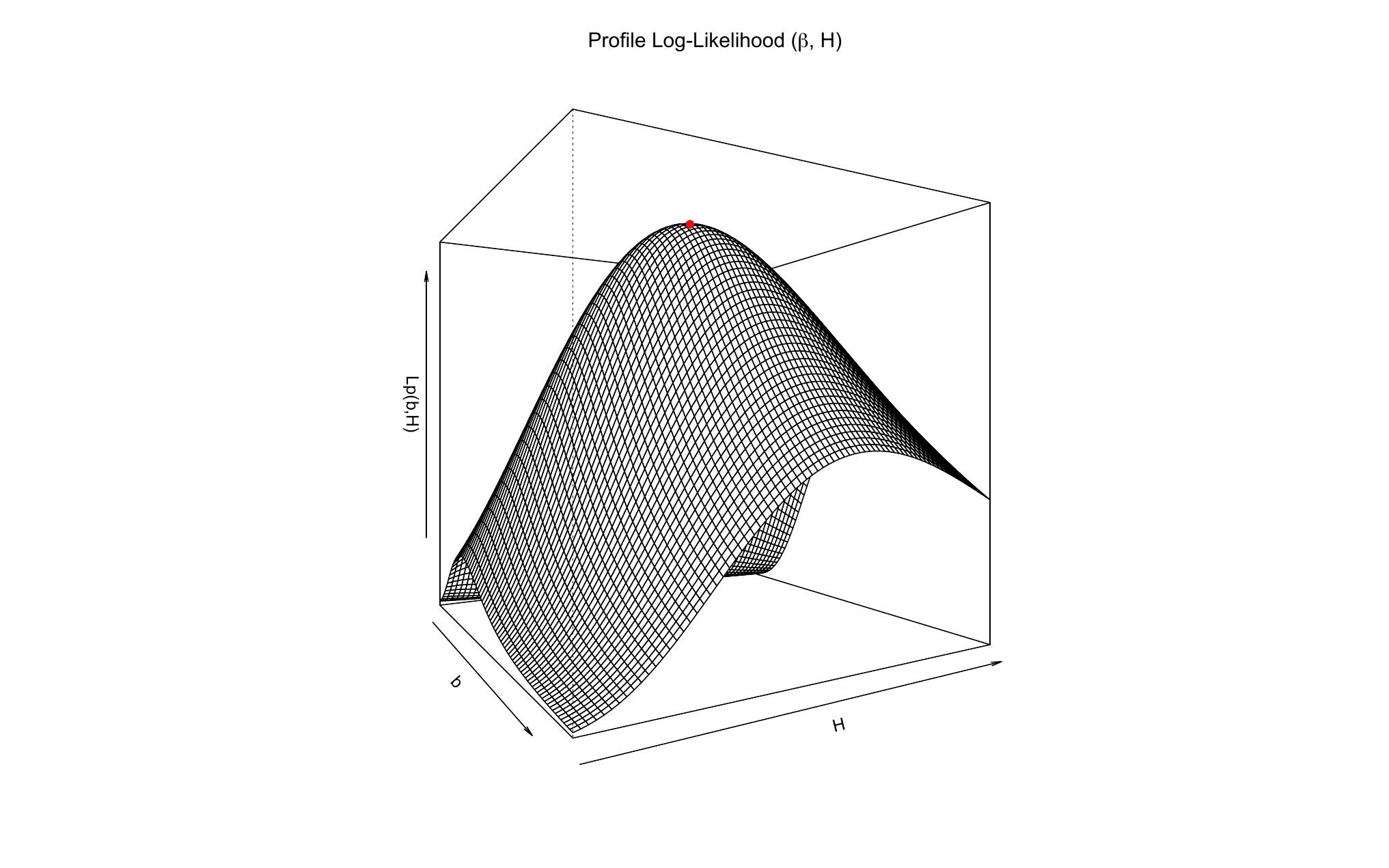}}\hspace{3mm}
\subfigure[latitude]{\includegraphics[width=63mm]{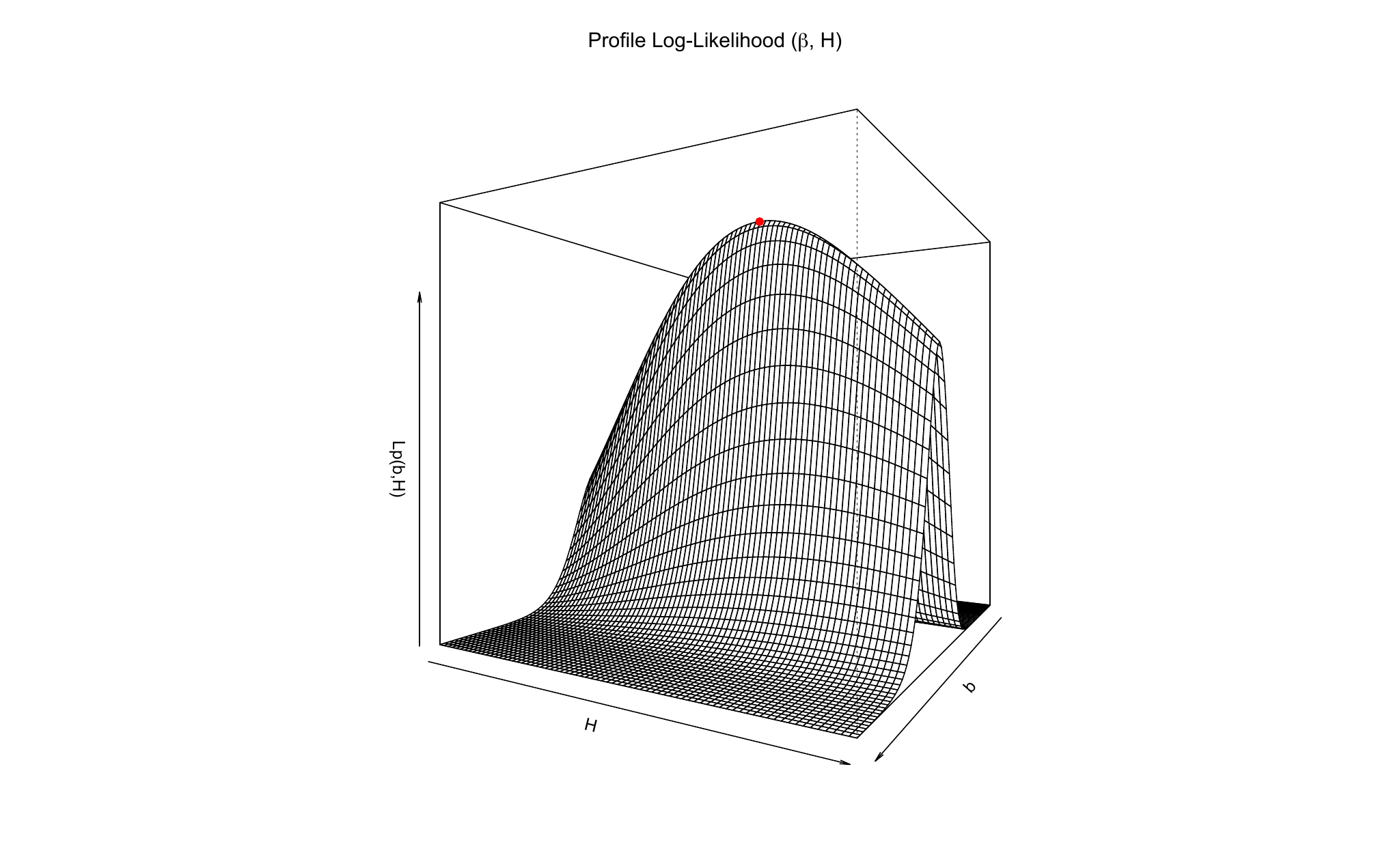}}
\caption{Profile Likelihood $(\beta,h)$ for longitude and latitude (Fin Whale $\#7$). The red points indicate the maximum value of the Log-likelihood $(\beta,H)$ function of each axis. } \label{LP3}
\end{figure}

\end{appendix}

\begin{acks}[Acknowledgments]

\end{acks}

\end{document}